\documentclass[aps,nofootinbib,prd,superscriptaddress,tightenlines,notitlepage,twocolumn,showpacs,floatfix]{revtex4-2}
\usepackage[colorlinks]{hyperref}
\usepackage{tabularx}
\usepackage{bm}
\usepackage{graphicx}
\usepackage{amssymb,bm,tensor}
\usepackage{xcolor}
\usepackage{amsmath}
\usepackage[varg]{txfonts}
\usepackage{enumerate}
\usepackage[normalem]{ulem}
\usepackage{bm}
\usepackage[OT1]{fontenc}
\usepackage{threeparttable}
\usepackage{mathtools}

\usepackage{scalerel,stackengine}
\stackMath
\newcommand\reallywidehat[1]{%
\savestack{\tmpbox}{\stretchto{%
  \scaleto{%
    \scalerel*[\widthof{\ensuremath{#1}}]{\kern-.6pt\bigwedge\kern-.6pt}%
    {\rule[-\textheight/2]{1ex}{\textheight}}
  }{\textheight}%
}{0.5ex}}%
\stackon[1pt]{#1}{\tmpbox}%
}

\begin{document}

\title{Quasi-Normal Mode Ringing of Binary Black Hole Mergers in Scalar-Gauss-Bonnet Gravity}

\date{\today}

\author{Zexin Hu}
    \email{huzexin@pku.edu.cn}
\affiliation{Department of Astronomy, School of Physics, Peking University, Beijing 100871, China}
\affiliation{Kavli Institute for Astronomy and Astrophysics, Peking University, Beijing 100871, China}
\affiliation{Theoretical Astrophysics, Eberhard Karls University of T\"ubingen. 72076 T\"ubingen, Germany}

\author{Daniela D. Doneva}
	\email{daniela.doneva@uni-tuebingen.de}
\affiliation{Theoretical Astrophysics, Eberhard Karls University of T\"ubingen, 72076 T\"ubingen, Germany}
\affiliation{INRNE - Bulgarian Academy of Sciences, 1784 Sofia, Bulgaria}

\author{Stoytcho S. Yazadjiev}
\affiliation{Department of Theoretical Physics, Sofia University ``St. Kliment Ohridski'', 5 J. Bourchier Blvd. Sofia 1164, Bulgaria}
\affiliation{Institute of Mathematics and Informatics, Bulgarian Academy of Sciences, Acad. G. Bonchev St. 8, Sofia 1113, Bulgaria}

\author{Lijing Shao}
\affiliation{Kavli Institute for Astronomy and Astrophysics, Peking University, Beijing 100871, China}
\affiliation{National Astronomical Observatories, Chinese Academy of Sciences, Beijing 100012, China}

\begin{abstract}

Observations of gravitational waves (GWs) generated by binary black hole (BBH) mergers provide us with a powerful way to explore the strong and highly dynamical regime of gravity theories. The ringdown of BBH merger, consisting of a series of quasi-normal modes (QNMs), is of particular interest for both the black hole (BH) spectroscopy and the inspiral-merger-ringdown consistency check. Unlike the QNM frequencies that only depend on the properties of the remnant BH, the excitation amplitudes and phases of QNMs depend on the progenitor system, and calculating them is beyond the perturbative approach. In this paper, by performing self-consistent fully non-linear simulations of BBH merger in shift-symmetric scalar-Gauss-Bonnet (sGB) gravity as well as in sGB gravity allowing for scalarization, and extracting the QNM excitation, we explore the possible deviations from GR at the ringdown stage. We numerically verify that the mode frequencies are consistent with the theory prediction, and provide the fitting results of mode amplitudes and phases. We find relatively small changes in the mode excitation, considering that the largest coupling we used in the simulations is close to the limit of loss of hyperbolicity. To demonstrate that our results are robust against the eccentricity caused by the imperfect initial data, we also perform eccentricity reduction and estimate the effect caused by the initial eccentricity. These studies are useful for understanding the ringdown in sGB gravity.

\end{abstract}

\maketitle


\allowdisplaybreaks

\section{Introduction}

The observations of gravitational waves (GWs) generated by compact object coalescence open a new area of GW astronomy and provide a powerful way to explore the strong and highly dynamical regime of gravity theory~\cite{LIGOScientific:2016aoc,LIGOScientific:2016lio,LIGOScientific:2021sio}. The final stage of the binary black hole (BBH) merger, known as the ringdown, which consists of a series of quasi-normal modes (QNMs)~\cite{Berti:2009kk}, can give unique tests of the black hole (BH) theory through the so-called BH spectroscopy~\cite{Berti:2025hly}. As described by the linear perturbation theory~\cite{Teukolsky:1972my,Teukolsky:1973ha}, the QNMs in the ringdown stage have mode frequencies that only depend on the properties of the final BH. In general relativity (GR), for an uncharged astrophysical BH, once its mass and spin are known, all its QNM frequencies are fixed~\cite{Leaver:1985ax,Onozawa:1996ux,Berti:2003jh}. Conversely, measuring the frequencies of several QNMs will provide an overdetermined system that gives a consistency check of the underlying theory~\cite{Isi:2019aib}.

The excitation of QNMs presented in the ringdown stage, however, depends on the progenitor system, and this dependence is beyond the scope of linear perturbation theory. Green's function techniques only imply that the QNM amplitudes can be represented by the product of excitation factors and initial data dependent integrals~\cite{Leaver:1986gd,Andersson:1995zk,Berti:2006wq,Silva:2024ffz}. Although there is a very recent study of extreme-mass-ratio inspiral into Schwarzschild BH that suggests the possibility of analytically modeling the inspiral imprint onto ringdown signals~\cite{DeAmicis:2025xuh}, the current understanding of ringdown excitation is mainly based on numerical relativity (NR) simulations. Fitting a large number of NR simulations can provide detailed mode excitation amplitudes and phases as functions of the parameters of the progenitor system~\cite{Buonanno:2006ui,Berti:2007fi,Cheung:2023vki}, which is important for the inspiral-merger-ringdown consistency check. Meanwhile, interesting quasi-universal relations between the excitation of linear and nonlinear modes are found~\cite{Cheung:2023vki,Zhu:2024rej}. Numerical waveforms also provide insight into the problem of the ringdown starting time, which is related to the nonlinear response at the end of the merger and is important in ringdown data analysis~\cite{Giesler:2019uxc}.

Although GR has passed all tests with flying colors so far~\cite{Will:2018bme}, it is believed that gravity needs to be quantized so that GR could only be an effective field theory (EFT) below some energy scale~\cite{Weinberg:2021exr}. Testing gravity then requires us to know not only the predictions of GR but also how things can deviate from GR in the unconstrained regime. Theories that can evade the no-hair theorem in GR have recently attracted a lot of attention~\cite{Herdeiro:2015waa,Doneva:2022ewd,Yazadjiev:2025ezx}. One well-studied example is the so-called scalar-Gauss-Bonnet (sGB) gravity theory, which can also be motivated as a low-energy EFT of a quantum theory of gravity with an additional scalar field coupled to gravity through the Gauss-Bonnet invariant. Depending on the coupling, sGB gravity can provide interesting phenomena such as spontaneous scalarization of BHs while coincident with GR in the weak field limit~\cite{Doneva:2017bvd,Silva:2017uqg,Antoniou:2017acq}. Thus, it is a concrete example for us to study the possible deviation from GR in the strong-field regime.

To study the ringdown of BBH mergers in sGB gravity, one needs to rely on NR simulations in the modified gravity theories, which is a rapidly developing area. One of the key requirements for performing the numerical evolution of modified gravity theories is to find well-posed formulations~\cite{Sarbach:2012pr}. Based on recent studies, which prove that Horndeski theories (including sGB gravity as a subclass) are well-posed in a modified generalized harmonic gauge and weak coupling limit~\cite{Kovacs:2020ywu,Kovacs:2020pns}, and also the later following studies that extend the results to singularity avoiding coordinates~\cite{AresteSalo:2022hua,AresteSalo:2023mmd}, numerical codes are built and are successfully used to perform $3+1$ evolution of BH systems ~\cite{East:2020hgw,East:2021bqk,Corman:2022xqg,AresteSalo:2023hcp,Doneva:2023oww,Corman:2024cdr,AresteSalo:2025sxc,Lara:2025kzj,Corman:2025wun} as well as neutron stars \cite{East:2022rqi,Corman:2024vlk}  in sGB gravity and its extensions \cite{Doneva:2024ntw}. The phenomena of dynamical descalarization and spin-induced dynamical scalarization during BH coalescence are explored~\cite{Elley:2022ept}. Although a stronger coupling will give a larger deviation from GR, the coupling strength in numerical simulations is limited by the loss of hyperbolicity, which happens when parts of the system enter the strongly coupled regime~\cite{Ripley:2019hxt,Doneva:2023oww,R:2022hlf}. This loss of hyperbolicity is considered to be physical and thus places an upper limit on the possible deviation.

In this paper, we study the ringdown stage in sGB gravity, focusing primarily on the shift-symmetric theory as well as the case admitting spontaneous scalarization. By numerically simulating a single perturbed BH, we verify the theoretically calculated polar and axial mode frequencies in sGB gravity, which, contrary to GR, are no longer degenerate and have different values~\cite{Blazquez-Salcedo:2024oek,Chung:2024ira,Chung:2024vaf,Khoo:2024agm}. We further perform numerical simulations of nearly equal-mass, quasicircular, non-spinning BBH mergers in sGB gravity with different coupling strengths, and compare the mode excitation with previous studies based on GR simulations~\cite{Cheung:2023vki,Boyle:2019kee}. By fine-tuning the theory parameters, we also give a rough estimation of the largest deviation in the ringdown stage for sGB theory that permits the phenomenon of spin-induced dynamical scalarization~\cite{Elley:2022ept,Doneva:2023oww}. 

The paper is organized as follows. In Sec.~\ref{sec:sGB} we give a brief introduction to the sGB theory that we study in this paper, and in Sec.~\ref{sec:NR} we describe the numerical scheme that we use to perform the simulations. Section~\ref{sec:extract procedure} lists the detailed procedure for QNM extraction. We present the single BH simulations and their QNM extraction results in Sec.~\ref{sec:single BH}. The results for BBH merger simulations in shift-symmetric sGB theory are shown in Sec.~\ref{sec:BBH}. In Sec.~\ref{sec:phi2} we discuss the ringdown of the BBH merger with spin-induced dynamical scalarization. Our conclusions are presented in Sec.~\ref{sec:conclusions}.

\section{Scalar-Gauss-Bonnet Theory of Gravity}\label{sec:sGB}

Although sGB gravity is widely studied in literature~\cite{Doneva:2017bvd,Silva:2017uqg,Antoniou:2017acq}, we briefly present its formalism here to define the notations used in this paper. The action for a general sGB theory can be written as
\begin{equation}
    S=\frac{1}{16\pi}\int{\rm d}^4x\sqrt{-g}\left[R-\frac{1}{2}\nabla_\mu \varphi\nabla^\mu\varphi-V(\varphi)+f(\varphi)\mathcal{R}_{\rm GB}^2\right]\,,
\end{equation}
where $R$ is the Ricci scalar of the metric $g_{\mu\nu}$. An additional scalar field $\varphi$ is coupled to the Gauss-Bonnet invariant $\mathcal{R}_{\rm GB}^2$ defined as
\begin{equation}
    \mathcal{R}_{\rm GR}^2=R^2-4R_{\mu\nu}R^{\mu\nu}+R_{\mu\nu\alpha\beta}R^{\mu\nu\alpha\beta}\,,
\end{equation}
with $R_{\mu\nu}$ and $R_{\mu\nu\alpha\beta}$ being the Ricci and Riemann tensors respectively. $V(\varphi)$ is the scalar field potential, which is set to zero in our simulations for simplicity. The vanishing $V(\varphi)$ gives a theory with a massless scalar field that allows for significant scalar dipole radiation~\cite{East:2021bqk}. The coupling between the scalar field and the Gauss-Bonnet invariant is controlled by the function $f(\varphi)$. 

The field equations for a general form of $V(\varphi)$ and $f(\varphi)$, can be obtained by varying the action with respect to the metric and scalar field, and they read~\cite{East:2020hgw,Doneva:2023oww}
\begin{eqnarray}
    R_{\mu\nu}-\frac{1}{2}g_{\mu\nu}R+\Gamma_{\mu\nu}&=&\frac{1}{2}\nabla_\mu\varphi\nabla_\nu\varphi-\frac{1}{4}g_{\mu\nu}\nabla_\alpha\varphi\nabla^\alpha\varphi\nonumber\\
    &&-\frac{1}{2}g_{\mu\nu}V(\varphi)\,,\\
    \nabla_\alpha\nabla^\alpha\varphi&=&\frac{{\rm d}V(\varphi)}{{\rm d}\varphi}-\frac{{\rm d}f(\varphi)}{{\rm d}\varphi}\mathcal{R}_{\rm GB}^2\,,\label{eq:field:phi}
\end{eqnarray}
with 
\begin{eqnarray}
    \Gamma_{\mu\nu}&=&-\frac{1}{2}R\Omega_{\mu\nu}-\Omega_\alpha^{\ \ \alpha}\left(R_{\mu\nu}-\frac{1}{2}Rg_{\mu\nu}\right)+2R_{\alpha(\mu}\Omega_{\nu)}^{\ \ \,\alpha}\nonumber\\
    &&-g_{\mu\nu}R^{\alpha\beta}\Omega_{\alpha\beta}+R^\beta_{\ \ \mu\alpha\nu}\Omega_\beta^{\ \ \alpha}\,,
\end{eqnarray}
where 
\begin{equation}
    \Omega_{\mu\nu}=4\nabla_\mu\nabla_\nu f(\varphi)\,.
\end{equation}
The notion $(\mu\nu)$ means symmetrization over the $\mu\nu$ indices.
 
There can be a wide range of choices for $f(\varphi)$ that result in very different behaviors of the theory~\cite{Sotiriou:2014pfa,Doneva:2017bvd,Silva:2017uqg,Antoniou:2017acq}. The simplest choice is
\begin{equation}
    f(\varphi)=\lambda\varphi\,,
\end{equation}
which gives the so-called shift-symmetric sGB theory, where all stationary BHs have a non-trivial scalar field. The strength of the coupling $\lambda$ is limited by the existence condition of regular BH solutions in this theory~\cite{Sotiriou:2014pfa} as well as the binary merger \cite{Lyu:2022gdr} and pulsar observations \cite{Yordanov:2024lfk}. The hyperbolicity requirement for numerical evolution in general puts further constraints on the coupling strength~\cite{East:2020hgw}.

Another widely studied choice involves quadratic coupling as $\lambda \varphi^2$, which allows for the phenomenon of spontaneous scalarization~\cite{Doneva:2017bvd,Silva:2017uqg,Antoniou:2017acq}. To stabilize the scalarized BH in this kind of theory, higher-order coupling terms like $\beta\varphi^4$ are needed~\cite{Silva:2018qhn,Macedo:2019sem,Minamitsuji:2019iwp}. In this work, we adopt the coupling function~\cite{Doneva:2023oww} 
\begin{equation}\label{eq:cop_phi2}
    f(\varphi)=\epsilon\frac{\lambda}{2\beta}\left(1-e^{-\beta\varphi^2}\right)\,,
\end{equation}
so that its Taylor expansion contains the quadratic coupling we need and $\epsilon=\mp 1$. With the above coupling, one can easily see from the field equations that GR solutions that have $\varphi=0$ are also valid solutions in this theory. However, for a certain region of the parameter space, they are not linearly stable. Instead, the scalarized black holes can be the stable and thermodynamically preferred solutions~\cite{Doneva:2022ewd}. Depending on the sign of the coupling, one can have the normal spontaneous scalarization or the spin-induced scalarization~\cite{Dima:2020yac,Herdeiro:2020wei,Berti:2020kgk}. 

\section{Numerical Implementation}\label{sec:NR}

As an example belonging to the more general Horndeski theories, it is proved that the sGB theory has a well-posed formulation in modified generalized harmonic gauge~\cite{Kovacs:2020ywu,Kovacs:2020pns} as well as in singularity-avoiding coordinates~\cite{AresteSalo:2022hua,AresteSalo:2023mmd}. We perform our numerical simulations using the \texttt{GRFolres}~\cite{AresteSalo:2023hcp} code, which is based on \texttt{GRChombo}~\cite{Andrade:2021rbd},  and implements the $3+1$ nonlinear evolution of sGB theory in vacuum. The code employs the modified CCZ4 formulation described in Refs.~\cite{AresteSalo:2022hua,AresteSalo:2023mmd}, which is a formalism that is strongly hyperbolic in the weak coupling limit. The formalism achieves its well-posedness by introducing two auxiliary Lorentzian metrics 
\begin{equation}
    \tilde{g}^{\mu\nu}=g^{\mu\nu}-a(x)n^\mu n^\nu\,,\quad \hat{g}^{\mu\nu}=g^{\mu\nu}-b(x)n^{\mu}n^{\nu}\,.
\end{equation}
Here $n^{\mu}$ is the unit timelike vector normal to the $t={\rm const.}$ hypersurfaces while $a(x)$ and $b(x)$ are arbitrary positive functions chosen in such a way that the null cones of these auxiliary metrics do not intersect and lie outside the null cone of $g^{\mu\nu}$~\cite{Kovacs:2020pns,Kovacs:2020ywu}. Viewed in the modified harmonic gauge, the auxiliary metrics break the degeneracy among the gauge-condition violating modes, pure gauge modes, and physical modes, which then provide a more stable structure of the principal symbol under perturbations. Furthermore, as in the usual CCZ4 formulation~\cite{Alic:2011gg}, additional damping terms with coefficients $\kappa_1>0$ and $\kappa_2>-2/(2+2b(x))$ ensure that the constraint violating modes are exponentially suppressed~\cite{AresteSalo:2023mmd}. In our simulations, we usually set $a(x)=0.2$ and $b(x)=0.4$, which is proven to lead to a stable evolution~\cite{East:2021bqk,Doneva:2023oww}, except that in dynamical scalarization we use $a(x)=0.1$ and $b(x)=0.2$. We use $\kappa_1=0.6/M$, $\kappa_2=-0.1$ for single BH simulations, $\kappa_1=1.0/M$, $\kappa_2=-0.1$ for BBH simulations, and $\kappa_1=0.6/M$, $\kappa_2=-0.15$ for dynamical scalarization. The Kreiss-Oliger numerical dissipation coefficient is set to $\sigma=0.5$. See Refs.~\cite{Clough:2015sqa,Radia:2021smk} for the detailed definition and implementation of these parameters in GR and in sGB gravity \cite{AresteSalo:2023mmd,Doneva:2023oww,AresteSalo:2025sxc}.

As studied in previous literature~\cite{Figueras:2020dzx,Figueras:2021abd,Doneva:2023oww}, when one gradually leaves the weak coupling limit, a nonhyperbolic region will inevitably develop first inside the BH horizon. As the simulation is performed in a horizon-penetrating puncture gauge, this area will lead to the breakdown of the code regardless of whether it is physically related to the outer-horizon problem. Thus, following previous studies~\cite{Figueras:2020dzx,Figueras:2021abd,AresteSalo:2022hua,AresteSalo:2023mmd,Doneva:2023oww}, we evade this problem by smoothly turning off the coupling $f(\varphi)$ inside the BH horizon. More specifically, we replace the coupling as
\begin{equation}\label{eq:cutoff}
    f(\varphi)\rightarrow\frac{f(\varphi)}{1+e^{-\beta_{\rm ex}(\chi-\chi_{\rm ex})}}\,,
\end{equation}
where $\chi$ is the usual variable introduced in conformal decomposition, namely $\chi=\det (\gamma_{ij})^{-1/3}$ with $\gamma_{ij}$ being the physical spatial metric. $\chi_{\rm ex}$ and $\beta_{\rm ex}$ are the parameters that mainly control where and how fast the coupling is turned off. In our simulations, we set $\beta_{\rm ex}=300$ and $\chi_{\rm ex}\sim0.1-0.18$ according to the coupling strength $\lambda$. We note that while the above replacement allows for a hyperbolic evolution, there will be a (sometimes large) Hamiltonian constraint violation inside the cutoff region. Clearly, this is caused mainly by the simple replacement 
\begin{equation}
    \frac{{\rm d}f(\varphi)}{{\rm d}\varphi}\rightarrow\frac{{\rm d}f(\varphi)/{\rm d}\varphi}{1+e^{-\beta_{\rm ex}(\chi-\chi_{\rm ex})}}\,,
\end{equation}
that violates the field equation. Nevertheless, as long as the constraint violation region is well inside the apparent horizon (AH), the physics of the system outside the horizon should be guaranteed. 

We describe the details of the initial data for our simulations in Sec.~\ref{sec:single BH} and Sec.~\ref{sec:BBH} as they are different for single BH and BBH cases.

\section{Quasinormal mode extraction procedure}\label{sec:extract procedure}

As we focus on the ringdown in the sGB theory, we perform QNM extraction based on the method discussed in Refs.~\cite{Cheung:2023vki,Mitman:2025hgy}. In this section, we summarize the procedure used for our QNM extraction. 

In our simulations, the GWs are extracted using a spin-weighted spherical harmonic decomposition of the Newman-Penrose scalar at a finite radius~\cite{Radia:2021smk}
\begin{equation}
    \Psi_4=\sum_{l,m}\prescript{}{-2}{Y}_{lm}\Psi_4^{lm}\,,
\end{equation}
where $\prescript{}{-2}{Y}_{lm}$ are the spin-weighted $s=-2$ spherical harmonics and $\Psi_4=\ddot{h}$ with $h$ being the GW strain. In the ringdown stage, the decomposition coefficients $\Psi_4^{lm}$ can be written as a superposition of QNMs 
\begin{equation}\label{eq:fitting ansatz}
    \Psi_4^{lm} (t)=\sum_{k=1}^N B_k e^{-i(\omega_{r,k}+i\omega_{i,k})(t-t_{\rm peak})-i\varphi_k}\,,
\end{equation}
where $\omega_{r,k}+i\omega_{i,k}$ is the complex QNM frequency and $B_k$ and $\varphi_k$ are the mode amplitude and phase at the reference time $t_{\rm peak}$. To compare with the previous study~\cite{Cheung:2023vki}, we choose $t_{\rm peak}$ to be the time of the maximum strain $|h|$ of the $lm=22$ multipole waveform. 
The commonly used mode amplitude and phase $A_k$, $\phi_k$ defined by 
\begin{equation}
    h_{lm} (t)=\sum_{k=1}^N A_k e^{-i(\omega_{r,k}+i\omega_{i,k})(t-t_{\rm peak})-i\phi_k}\,,
\end{equation}
are related to $B_k$ and $\varphi_k$ via
\begin{equation}
    B_ke^{-i\varphi_k}=-(\omega_{r,k}+i\omega_{i,k})^2A_ke^{-i\phi_k}\,.
\end{equation}

For extracting the QNMs, we fit the $\Psi_4^{lm}$ extracted from the outermost reliable finite radius in our simulations, $r=90\,M$, where $M$ is the mass unit used in the simulations. Although we used a much larger computational domain for the simulations, considering the balance of computation cost and the refinement level needed for extraction, we use this relatively small extraction radius. In principle, an extrapolation procedure can be applied to obtain a waveform at the future null infinity~\cite{Boyle:2019kee}. However, it only introduces a small change in the fitting results compared to the uncertainties of the fitting procedure. Also, considering the unknown errors introduced in the extrapolation procedure~\cite{Mitman:2020bjf}, we choose to use the finite radius waveform. Future simulations with the Cauchy characteristic extraction method can provide more reliable waveforms at the future null infinity~\cite{Bishop:1996gt,Taylor:2013zia,Giannakopoulos:2023zzm}.

As discussed in Ref.~\cite{Cheung:2023vki}, although we expect that $\Psi_{4}^{lm}$ is dominated by the natural linear mode $lm0$ that has the same $l$ and $m$ numbers as the underlying harmonic, due to various effects~\cite{Baibhav:2023clw,Berti:2014fga,London:2014cma,Dhani:2020nik,Mitman:2022kwt}, there will be other modes or even not mode components presented in $\Psi_{4}^{lm}$. Therefore, the total mode number $N$ in Eq.~(\ref{eq:fitting ansatz}) also needs to be determined for each fitting. In general, a larger number of $N$ will lead to a more stable fitting result for the dominant modes according to the stability criteria defined later~\cite{Cheung:2023vki}. However, fitting with too many modes might overfit the waveform and also lead to the numerical problem of finding a global minimum in a high-dimensional parameter space. We state our choice of $N$ for each fitting result in the sections below. 

When fitting with Eq.~(\ref{eq:fitting ansatz}), the modes included in the fitting can be further separated into two kinds, $N=N_{\rm free}+N_{\rm fix}$. For the kind we call free mode, we perform frequency-agnostic fitting, in which we treat both the mode frequency $\omega_{r,k}$, $\omega_{i,k}$, and the mode amplitude and phase $B_k$, $\varphi_k$ as free parameters. For the other kind, we call fixed mode, we fix the mode frequency and only fit for its amplitude and phase. Unlike in GR, the QNM frequencies in sGB theory for BH with arbitrary spin have only been calculated very recently and are limited to a small number of modes~\cite{Chung:2024ira,Chung:2024vaf,Xiong:2024urw,Khoo:2024agm}. Although fixing mode frequency in general gives better results in amplitude and phase extraction, for most modes, we have to perform frequency-agnostic fitting. 

Following Refs.~\cite{Cheung:2023vki,Mitman:2025hgy}, for each $\Psi_4^{lm}$, we perform a series of fittings with different starting time $t_0\in[t_{\rm peak},t_{\rm peak}+50\,M]$, {while the end time $t_{\rm end}$ is fixed. Based on the waveform quality, we use $t_{\rm end}=t_{\rm peak}+100\,M$ for single BH perturbation simulations and $t_{\rm end}=t_{\rm peak}+130\,M$ for BBH simulations, after which a clear numerical noise can be seen in the waveform. 
The presence of a QNM in the waveform is verified by the stability criteria~\cite{Cheung:2023vki,Mitman:2025hgy} that the extracted mode frequency, as well as its amplitude and phase, should be stable at least for some time window of the starting time. We define the instability of a fixed mode in a given time window to be 
\begin{equation}
    \sigma_{\rm fix}=\sqrt{\left(\frac{\Delta {\rm Re}\,C_{\rm QNM}}{|C_{\rm QNM}|}\right)^2+\left(\frac{\Delta {\rm Im}\,C_{\rm QNM}}{|C_{\rm QNM}|}\right)^2}\,,
\end{equation}
where $C_{\rm QNM}=B_ke^{-i\varphi_k}$ is the complex amplitude of the mode, and $\Delta$ represents the standard deviation. For free mode, we further consider the variation in the mode frequency as
\begin{equation}
    \sigma_{\rm free}=\sqrt{\sigma_f^2+\frac{(\Delta \omega_r)^2+(\Delta \omega_i)^2}{\omega_r^2+\omega_i^2}}\,.
\end{equation}
The above definition of instability depends on the length of the considered time window. Considering the damping nature of the QNMs, we adopt the method proposed in Ref.~\cite{Mitman:2025hgy} where the length of the time window is set to be proportional to the mode damping time. More precisely, for mode $lmn$, we set the length of the time window to be 
\begin{equation}
    \tau_{lmn}=-\frac{\ln 10}{\omega_{i,\,lmn,\,{\rm GR}}}\,,
\end{equation}
where the subscript ``GR'' denotes that we use the GR mode frequency to estimate the mode damping time. That is, we use the mode frequency of a Kerr BH with the same mass and spin as the remnant BH in the simulation, as for most modes we do not have the theoretically calculated frequencies.

Once a mode is confirmed to exist, that is, its instability is smaller than the given criterion $\sigma_{\rm fix/free}<0.2$ for at least one time window, we obtain the final fitting result for the mode parameters as the average value in the most stable time window. To extract $lmn$ mode, we always fit $\Psi_4^{lm}$ with the same $l$ and $m$ values.

\section{Single BH perturbation}\label{sec:single BH}

In this section, we present the simulations and extraction results of a single perturbed BH in shift-symmetric sGB theory. We excite polar and axial modes separately and show that their QNM frequencies are consistent with theoretical expectations. We also discuss the difficulties of extracting polar and axial modes simultaneously when they are both present. 

For the single perturbed BH simulations presented in this section, we set the size of the computational domain to be $L=512\,M$ along each of the coordinate directions, and we use $7$ refinement levels with a refinement ratio of $2:1$. The coarsest level is set to have $\Delta x_{\rm c}=L/192$ so that the finest resolution is $\Delta x=\Delta x_c/2^7=M/48$.

Our initial data are isolated Kerr BHs in quasi-isotropic coordinates~\cite{Liu:2009al} with initial angular momentum $a_0/M_0=0.6$ and initial mass $M_0/M=1$. The coupling constant in shift-symmetric sGB theory is chosen to be $\lambda/M^2=0.18$, which is close to the largest coupling that we can have for a stable hyperbolic evolution. On top of the GR background, we add a metric perturbation to excite the desired QNMs~\cite{Zhu:2024rej}. The perturbation is added in the angular sector of the metric and extrinsic curvature as 
\begin{eqnarray}
    g_{ij}&=&g_{ij}^{B}+h_{ij}^p+h_{ij}^a\,,\\
    K_{ij}&=&K_{ij}^B-\frac{1}{2}\partial_t h_{ij}^p-\frac{1}{2}\partial_t h_{ij}^a\,,
\end{eqnarray}
where $g_{ij}^B$ and $K_{ij}^B$ are the Kerr background  metric and extrinsic curvature. $p$ and $a$ denote the polar and axial parts of the perturbation, so that
\begin{eqnarray}
    h_{ij}^p&=&\sum_{lm}A_{lm}^pR_{lm}^p(r)Y^{lm}_{ij}(\theta,\phi)\,,\\
    h_{ij}^a&=&\sum_{lm}A_{lm}^aR_{lm}^a(r)X^{lm}_{ij}(\theta,\phi)\,.
\end{eqnarray}
$Y^{lm}_{ij}$ and $X^{lm}_{ij}$ are the even and old parity tensor spherical harmonics. We follow the convention in Ref.~\cite{Martel:2005ir} and set $Y^{lm}_{ri}=X^{lm}_{ri}=0$ for $i=\{r,\theta,\phi\}$. The radial profile $R_{lm}^{p/a}(r)$ is chosen to be the same for different $lm$ and parity as 
\begin{equation}
    R_{lm}^{p}(r)=R_{lm}^a(r)=e^{-\frac{(r-r_0)^2}{w^2}}\,,
\end{equation}
where $r_0$ is set to $50\,M$ and the width of the radial profile $w$ is chosen to be $1\,M$ for a better excitation of the corotating modes~\cite{Zhu:2024rej}. 

The terms added to the extrinsic curvature aim to provide an ingoing trend of the initial perturbation. The time derivative of $h_{ij}$ should be understood as (consider an ingoing wave)
\begin{equation}
    \partial_t h_{ij}^p=\frac{{\rm d}R_{lm}^p(r)}{{\rm d}r}Y_{ij}^{lm}\,,
\end{equation}
and the same for the axial perturbation.

\begin{figure}[htbp]
  \centering
  \includegraphics[width=0.45\textwidth]{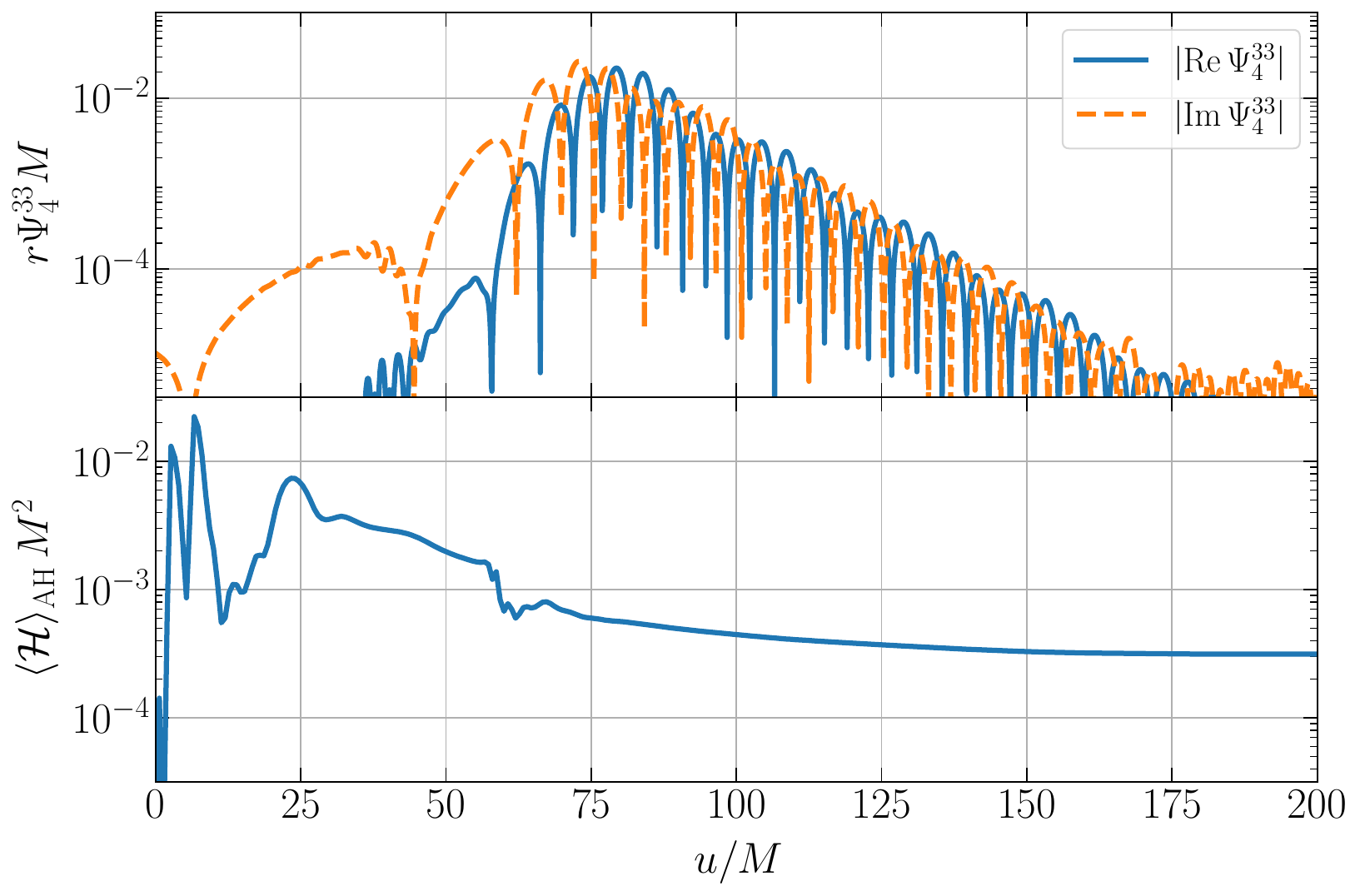}
  \caption{The waveform of $\Psi^{33}_4$ extracted at $r=90\, M$ and the average Hamiltonian constraint violation at the AH for a single perturbed BH simulation in shift-symmetric sGB gravity with an initial axial $l=3$, $m=3$ perturbation. $u=t-r$ is the Bondi time, such that one can roughly compare the quantities in the two panels. It can be seen that, when the ringdown starts, the constraint violation is largely damped.\label{fig:33_wave_Ham}}
\end{figure}

Figure~\ref{fig:33_wave_Ham} shows an example of a simulation with axial $l=3$, $m=3$ perturbation, that is, we only set $A_{33}^a$ to have a nonzero value. The upper panel of the figure shows the real and imaginary parts of $\Psi_4^{33}$ extracted at $r=90\, M$ and the lower panel shows the average Hamiltonian constraint violation at the AH.
Note that the initial perturbations added as above do not satisfy the constraint equations. Simply adding such a perturbation instead of using the conformal-thin-sandwich method~\cite{Pfeiffer:2004qz} to construct better initial data is mainly based on the following considerations. The code we used to evolve the space-time and the scalar field employs the modified CCZ4 formulation~\cite{Alic:2011gg,Alic:2013xsa,AresteSalo:2022hua,AresteSalo:2023mmd}, which is a constraint-violation damping formalism where the initial violation of the constraint equations can be damped, as shown in the lower panel of Fig.~\ref{fig:33_wave_Ham}. Furthermore, the background Kerr metric in GR is not a stable BH solution in the shift-symmetric sGB theory, although it satisfied the constraint equations at the beginning~\cite{East:2020hgw}. This will lead to a violent growth of the scalar field, causing an increase in the constraint violation once the simulation begins. Also, the initial data based on the quasi-isotropic coordinates need a period of $\sim10-20\,M$ to settle to a nearly stationary state due to the use of the puncture gauge~\cite{Doneva:2023oww}, which also leads to an increase of the constraint violation at the beginning as shown in the figure. Thus, although the initial perturbation we added is another contribution to the constraint violation, what we required is that, at the ringdown stage, the constraint violation is damped. In Fig.~\ref{fig:33_wave_Ham} we plot the waveform as a function of $u=t-r$, where $u$ is the Bondi time. It can be seen that when the ringdown is excited around the BH, the Hamiltonian constraint violation around the BH is largely damped, which provides a validation of our method. Nevertheless, carefully constructed initial data that take into account the scalar field and constraint-satisfied perturbation can be helpful, and should be further explored in future studies~\cite{Nee:2024bur,Brady:2023dgu}.

\begin{figure}[htbp]
  \centering
  \includegraphics[width=0.45\textwidth]{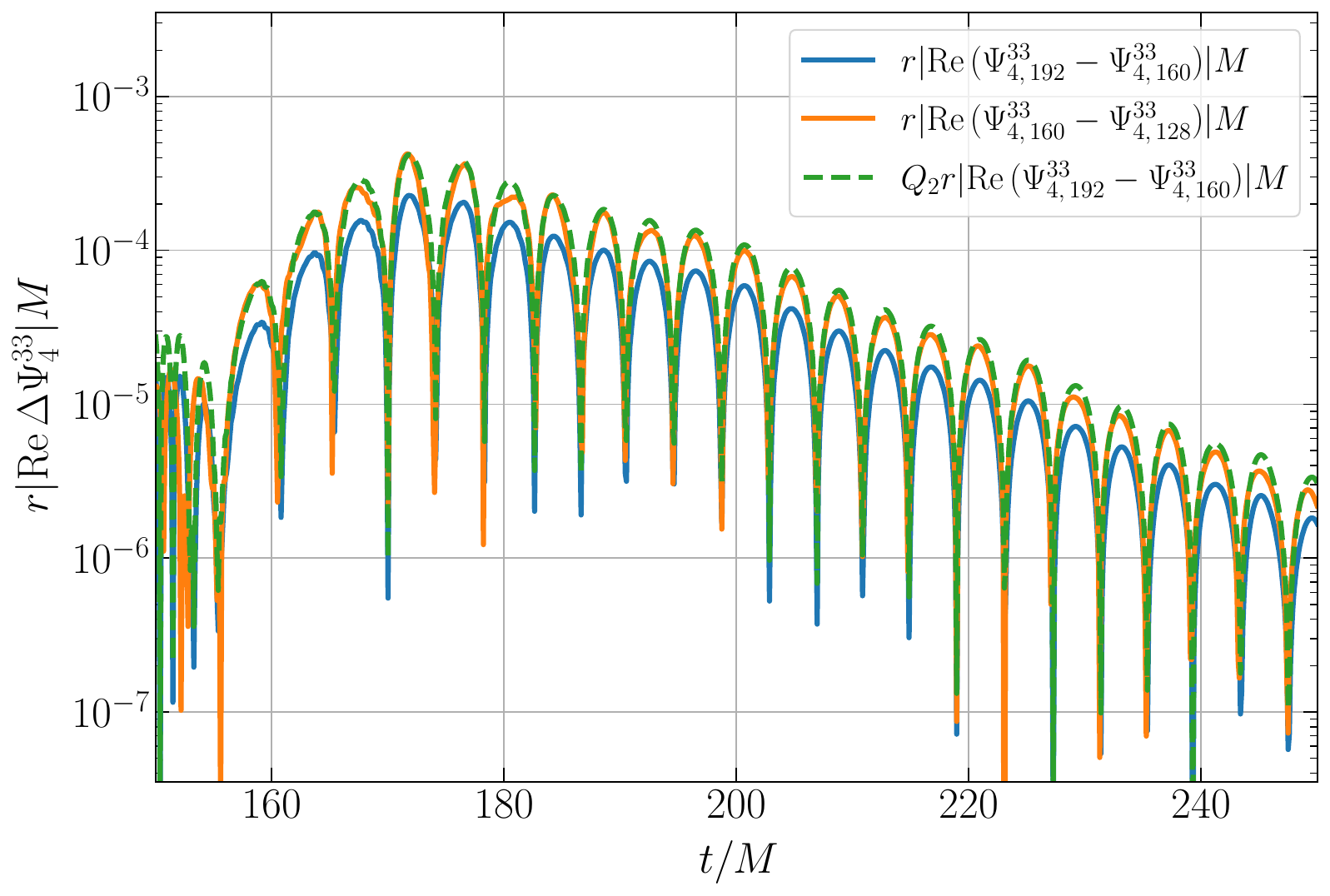}
  \caption{Convergence of the waveform ${\rm Re}\,\Psi_4^{33}$ for simulations of a single perturbed BH. The three simulations have finest grid resolutions $\Delta x=M/32$, $M/40$, and $M/48$, and they are denoted by the subscripts $128$, $160$, and $192$, respectively. The multiplying coefficient corresponding to a second-order convergence is $Q_2=(1/128^2-1/160^2)/(1/160^2-1/192^2)$. The results show that a second-order convergence is observed. The figure only shows the time window that has a significant ringdown signal. \label{fig:33_convergence}}
\end{figure}

We also perform convergence tests for our simulations. An example of the convergence of the waveform ${\rm Re}\,\Psi_4^{33}$ is shown in Fig.~\ref{fig:33_convergence}. The finest grid resolutions in the three simulations denoted by the subscripts $192$, $160$, and $128$ are $\Delta x=M/48$, $M/40$, and $M/32$, respectively. The figure only presents a time range where the ringdown signal dominates the waveform. It is clear that the waveform shows a second-order convergence that is consistent with the convergence test of the \texttt{GRChombo} code~\cite{Radia:2021smk}. We directly show the convergence of the waveform instead of the commonly used amplitude and phase~\cite{Radia:2021smk} as we focus on the ringdown stage, while smoothly evolving amplitude and phase are more properly defined for the inspiral.

\begin{figure}[htbp]
  \centering
  \includegraphics[width=0.45\textwidth]{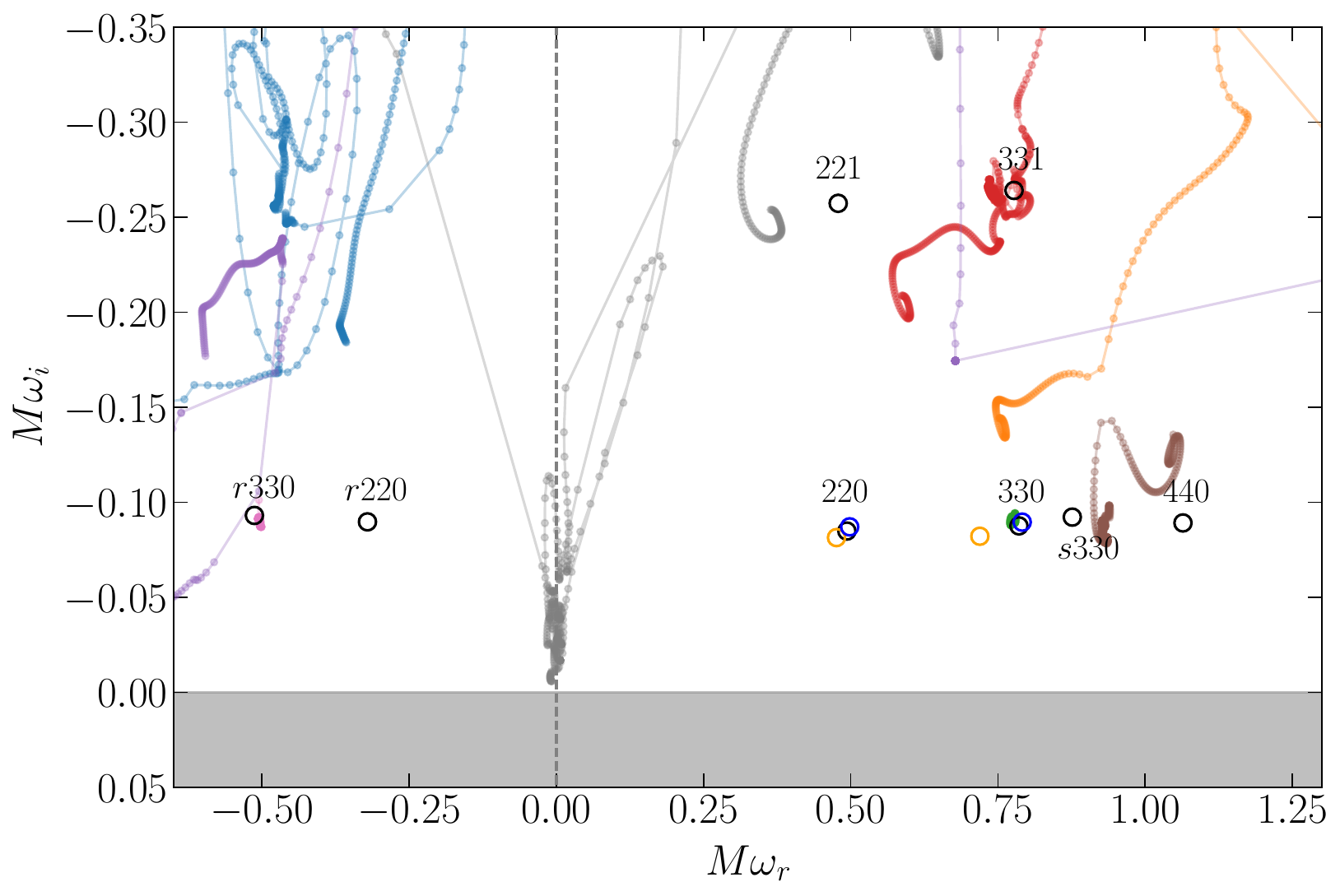}
  \caption{Fitting results of the QNM frequencies for $\Psi_4^{33}$ of the single perturbed BH simulation that has an initial axial $l=3$, $m=3$ perturbation. The dots with the same color show the evolution of the fitting result of a single mode. The black circles denote the tensorial mode frequencies of a Kerr BH, except for $s330$ which denotes the scalar-led $330$ mode in the test-field limit. The blue and orange circles show the axial and polar modes in the shift-symmetric sGB theory. For this figure, we use $N=8$. \label{fig:A_33}}
\end{figure}

\begin{figure}[htbp]
  \centering
  \includegraphics[width=0.45\textwidth]{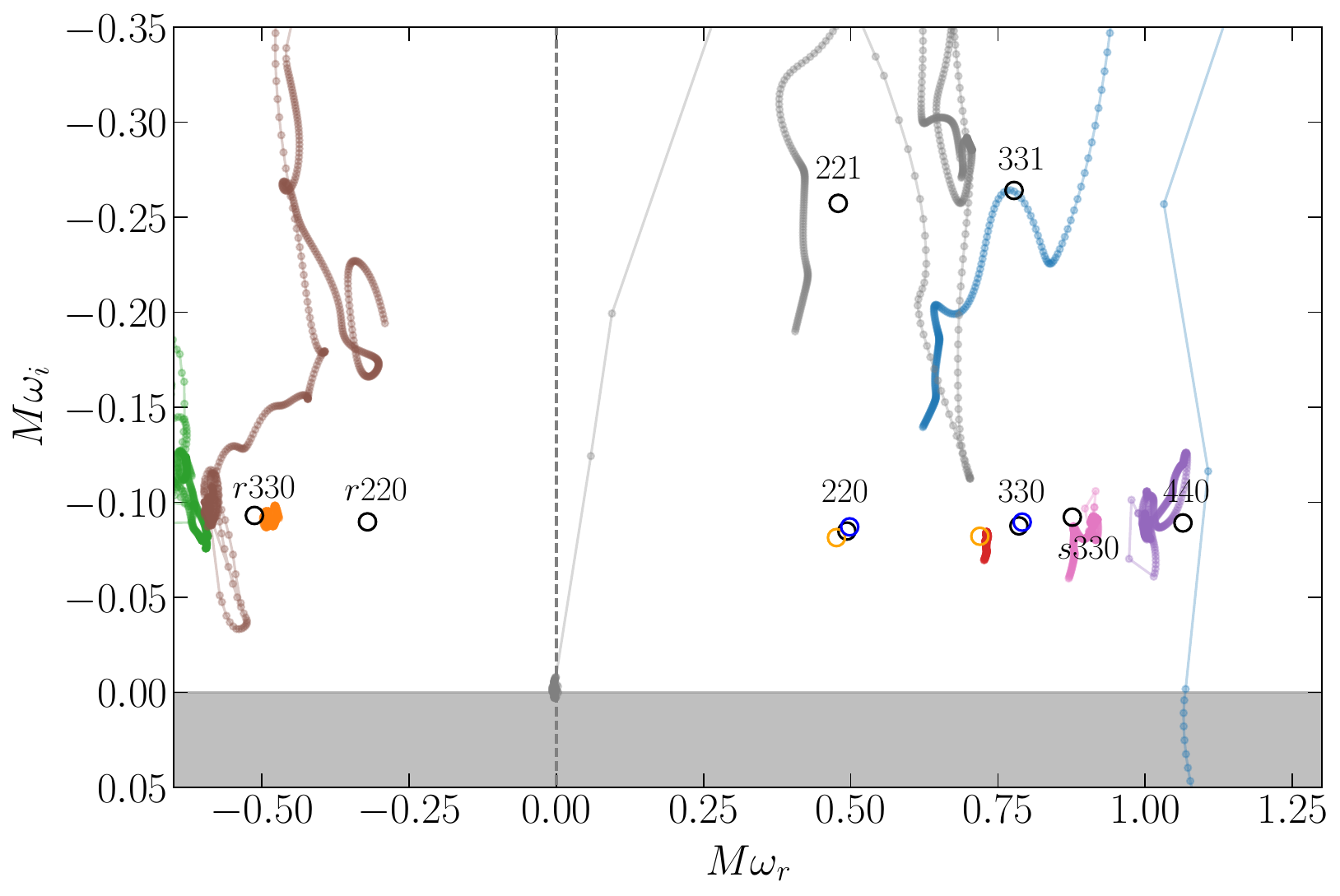}
  \caption{Similar figure as Fig.~\ref{fig:A_33} but with an initial polar $l=3$, $m=3$ perturbation. It is clear that at this time, the excited $330$ mode has a frequency that is consistent with the theoretical predictions for the polar mode.\label{fig:P_33}}
\end{figure}

In Fig.~\ref{fig:A_33} and Fig.~\ref{fig:P_33}, we show the frequency-agnostic extraction results of two simulations. The simulation for Fig.~\ref{fig:A_33} has initial data that contain axial $l=3$, $m=3$ perturbation. In the figure, we use black circles to denote the mode frequencies of a Kerr BH, $\omega_{\rm GR}(M_f,a_f)$, where $M_f$ and $a_f$ are the final mass and spin of the BH. The numbers near each circle denote the $lmn$ of the mode, while an $r$ before the number means retrograde mode, and $s$ denotes the scalar-led mode. For the scalar-led mode, we only show its test-field limit in the Kerr spacetime. The orange and blue circles show the corresponding polar and axial mode frequencies, $\omega_{p/a}(M_f,a_f,\lambda)$, which are also functions of the coupling constant $\lambda$. We only plot orange and blue circles for the $220$ and $330$ modes as they have been calculated and provided in tabulated form in Ref.~\cite{Chung:2024vaf}. The dots with the same color show the evolution of the fitting result of a single mode as the starting time changes.

From the two figures, one can clearly see that there are excitations of the $330$ mode (gray dots in the figure) and the retrograde $330$ mode. For the $330$ mode, the frequency is close to the predicted axial mode frequency in the shift-symmetric sGB theory. Note that we use the AH mass and spin to estimate the mode frequency, which might be biased due to the large coupling. A proper calculation should use the ADM mass and spin of the BH, which are different due to the presence of the scalar field~\cite{Chung:2024vaf}. Nevertheless, the difference between the polar and axial mode frequencies is large enough for us to distinguish the excitation of the two modes. To have a clean fitting for the desired mode, we use $N=8$. Figure~\ref{fig:P_33} is very similar to Fig.~\ref{fig:A_33} with the only difference that the initial perturbation is polar with $l=3$, $m=3$. One can clearly see that the excited $330$ mode is close to the polar frequency from Ref.~\cite{Chung:2024vaf}.

In both figures, one can also see that there is possibly a mode component that is close to the scalar-led 330 mode~\cite{Berti:2009kk}. Note that we only show the scalar-led mode frequency at the test-field limit for a Kerr BH, as there is a lack of calculation for the shift-symmetric sGB theory~\cite{Blazquez-Salcedo:2024oek}. The frequency change of the scalar-led mode, though, should be of the same order as the tensor mode. Recent studies suggest that accurately modeling the scalar-led modes is also essential for robust tests of GR in the ringdown regime~\cite{Crescimbeni:2024sam,Crescimbeni:2025kxi}.

\begin{figure*}[htbp]
  \centering
  \includegraphics[width=0.9\textwidth]{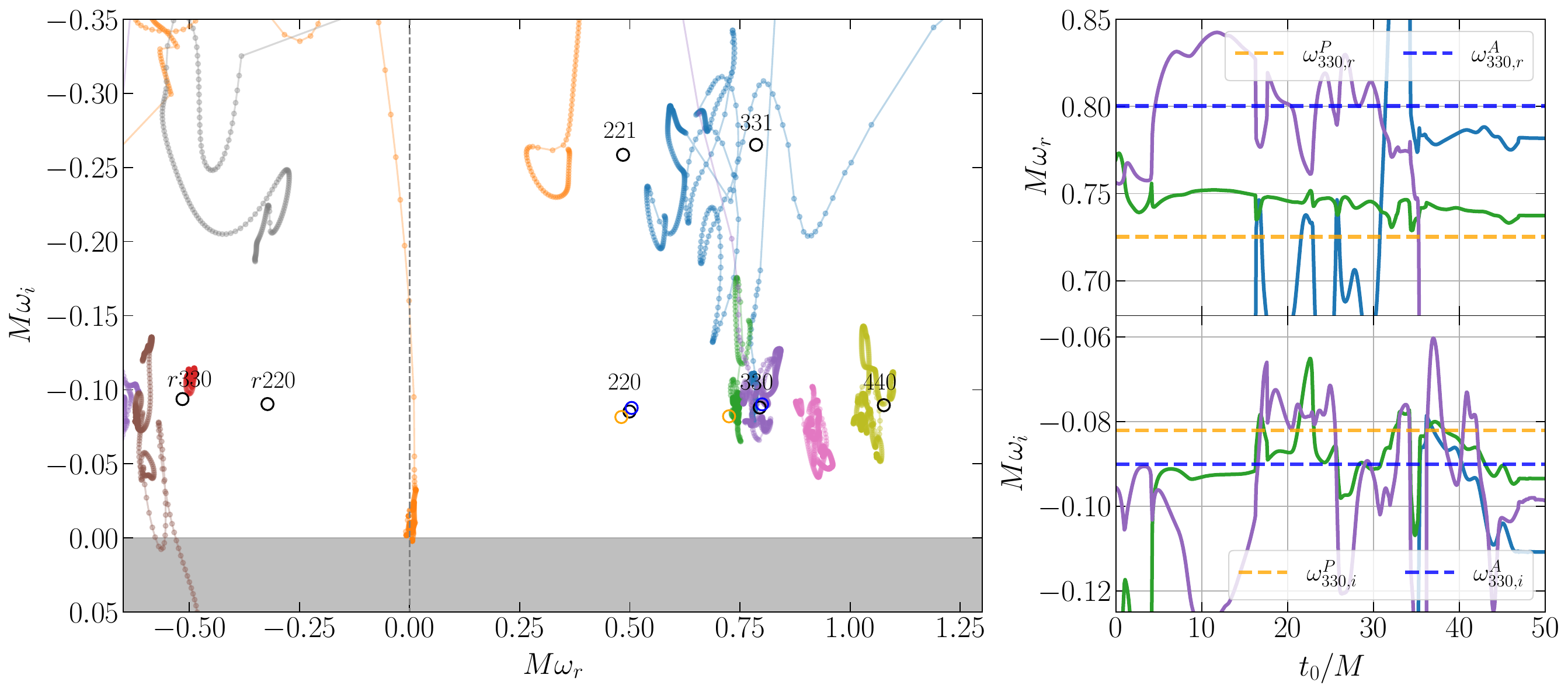}
  \caption{Similar figure as Fig.~\ref{fig:A_33} but with an initial $l=3$, $m=3$ perturbation that contains both polar and axial parts. In the right panel, we show the fitting result of the real and imaginary parts of the mode frequencies as a function of the fitting starting time $t_0$. The dashed lines show the estimation of the mode frequencies. Compared to Fig.~\ref{fig:A_33} and Fig.~\ref{fig:P_33} that contain only one mode centering at the estimated frequency, this figure shows a significantly more complex fitting result. Considering the additional mode, we use $N=9$ for the fitting. \label{fig:PA_33}}
\end{figure*}

As mentioned before, the coupling we used here is close to the largest coupling that still allows for a hyperbolic evolution. Also, according to Ref.~\cite{Chung:2024vaf}, for the BH spin we considered, among the modes with calculated frequencies, the $330$ mode gives the largest frequency difference between the polar and axial modes. However, even with such a large coupling, it is still difficult to extract polar and axial modes simultaneously based on the current simulation quality and the extraction procedure. 
In Fig.~\ref{fig:PA_33}, we show the fitting result for a simulation with an initial $l=3$, $m=3$ perturbation, where the polar and axial perturbations have the same amplitude. Compared to the fitting results of previous simulations that only have a single perturbation, the fitting results here are messier. In the right panel of the figure, we show the fitted mode frequencies as functions of the starting time $t_0$ of the fitting. Although the fitting seems to suggest that there are two modes close to the $330$ mode, the stability of the fitting is clearly worse than previous results. We have also checked that changing $N$ does not provide much better results. Considering the additional mode, we use $N=9$ here.

Frequency-agnostic fitting for two QNMs with very close frequencies is problematic in the following sense. Consider the superposition of the polar and axial modes that have similar frequencies, but with a Taylor expansion in time,
\begin{equation}\label{eq:expan1}
    Ae^{-i\omega_1t}+Be^{-i\omega_2t}=e^{-i\omega_{\rm GR} t}\left[A(1-i\tilde{\omega}_1t)+B(1-i\tilde{\omega}_2t)+\cdots\right]\,,
\end{equation}
where $\omega_{\rm GR}$ denotes the corresponding mode frequency in GR. $A$, $B$, $\omega_1$, and $\omega_2$ are complex numbers, and $\tilde{\omega}_1=\omega_1-\omega_{\rm GR}$, $\tilde{\omega}_2=\omega_2-\omega_{\rm GR}$ are small. One can rearrange the above expansion by collecting the terms that have the same order of $t$ as
\begin{equation}
     Ae^{-i\omega_1t}+Be^{-i\omega_2t}=e^{-i\omega_{\rm GR} t}\left[C_0+C_1t+C_2 t^2+C_3 t^3+\cdots\right]\,,
\end{equation}
where $C_i$ are also complex numbers that are determined by the  expansion~(\ref{eq:expan1}). For example, $C_0=A+B$. Fitting for $A$, $B$, $\omega_1$ and $\omega_2$ then is equivalent to fitting the coefficients $C_0$, $C_1$, $C_2$ and $C_3$. However, one may rewrite the above expansion as
\begin{equation}
   Ae^{-i\omega_1t}+Be^{-i\omega_2t}= e^{-i\omega_{\rm GR} t}\left[C_0e^{C_1 t}+O(t^2)\right]\,.
\end{equation}
 Therefore, the above formula suggests that, to confirm the existence of the two modes, one should at least fit out the coefficient of the $t^2$ term; otherwise, it is equivalent to a single mode. However, roughly speaking, extracting this coefficient requires a longer time window, which is not achievable for QNMs that have an exponentially decaying nature controlled by $e^{-i\omega_{\rm GR} t}$. Numerical errors will easily dominate the QNM before the $t^2$ term becomes important.

The situation will be even worse for the physically relevant cases where the ringdown is excited by nearly equal-mass BBH mergers. The largest coupling allowed in a BBH merger is limited by the small mass BH before the merger since the dimensionless coupling $\lambda/m_i^2$ needs to be smaller than the critical value for each BH~\cite{Sotiriou:2014pfa}. However, the dimensionless frequency difference $M\delta\omega$ of the polar and axial modes is proportional to $\lambda^2/M_f^4$~\cite{Chung:2024vaf}. For the best case with $M_f\sim2m_i$, the frequency difference will still be about $16$ times smaller than what we showed in this section. Distinguishing and extracting them in simulations as well as in real observations can be challenging.

\section{Ringdown of BBH merger in shift symmetric sGB gravity}\label{sec:BBH}

In this section, we explore the ringdown of nearly equal-mass, quasicircular, non-spinning BBH mergers in shift-symmetric sGB theory. We construct a series of simulations with the coupling constant $\lambda/M^2$ ranging from $0$ to $0.05$. We perform detailed QNM extraction to show the change in the QNM excitation caused by the scalar coupling.

\subsection{Numerical setup}

To study the QNM excitations of the BBH merger in shift-symmetric sGB theory, we perform simulations of nearly equal-mass, quasicircular, non-spinning BBH merger simulations. We set the mass ratio $q=1.2$ with the two initial BHs having mass $m_1=0.6\, M$ and $m_2=0.5\,M$. We denote the total mass as $M_t=m_1+m_2=1.1\,M$. We do not use the equal-mass ratio as we also want to see the excitation of the modes besides the dominant $220$ mode. On the other hand, as discussed before, the largest coupling allowing for hyperbolic evolution, so as the largest deviation from GR, is limited by the mass of the smaller BH. A mass ratio $1.2$ thus provides a relatively large excitation of, for example, the $330$ mode, while it can support a relatively large coupling for the remnant BH. The initial separation of the two BHs is set to $d=12\,M$ and the coupling constant $\lambda$ ranges from $0$ to $0.05\,M^2$. We use a computational domain with $L=1024\,M$ along each of the coordinate directions, and we apply the reflection symmetry about the orbital plane, which reduces half of the domain. Due to computational cost, we use $9$ refinement levels and the coarsest level is set to have $\Delta x_c=L/128$. With a refinement ratio of $2:1$, the finest resolution is $\Delta x=\Delta x_c/2^9=M/64$, which provides more than $60$ grid points across the AH of the small BH.

The initial data for BBH are generated by the \texttt{TwoPunctures} solver integrated in the \texttt{GRChombo}, which creates Bowen-York initial data for two-puncture BHs using a single domain spectral method~\cite{Ansorg:2004ds}. We calculate the initial momenta of the punctures for a quasicircular orbit with the 3PN approximation described in Refs.~\cite{Healy:2017zqj,Brown:2007jx}. We shall note that, as for single BH simulations, the initial data of the BBH merger are constructed fully in GR without a scalar field. Although it satisfies the constraint equations, the violent growth of the scalar field at the beginning will disturb the system and introduce eccentricity. Therefore, even though the method in Ref.~\cite{Healy:2017zqj} provides rather low eccentricity for GR simulations (about $10^{-3}$), we find that the eccentricity of our simulations with large coupling is higher (at the level of $0.04$ for the largest value of $\lambda$). 

\subsection{Effect of eccentricity}\label{subsec:ecc}

As we want to quantitatively study how the scalar coupling affects the QNM excitation, it is important to estimate and control the influence of the eccentricity. Although it might be believed that a small initial eccentricity will have negligible effects on the ringdown stage due to the fast circularization in the late inspiral~\cite{Sperhake:2007gu,Hinder:2007qu}, a proper comparison based on simulation is now possible. Here, we give a simple estimation of the effects caused by the eccentricity on QNM excitation based on GR simulations. We select nine non-spinning BBH merger simulations from the SXS catalog~\cite{Boyle:2019kee} with eccentricity ranging from $0$ to $0.1$. The mass ratio is chosen to be $q=2$ so that we can have enough simulations and can also extract modes besides the dominant $220$ mode. The simulation numbers we use are the following: $1222$, $2497$, $2425$, $1167$, $1166$, $1164$, $1165$, $1364$, $1365$, and they are ordered by their reference eccentricities. We extract the $220$, $330$, and $210$ mode amplitudes using the public python code \texttt{Jaxqualin}~\cite{Cheung:2023vki}, which employs a similar extraction method as we described in Sec.~\ref{sec:extract procedure} but takes advantage of assuming GR and adjusts to the SXS catalog. In Fig.~\ref{fig:sxs_q2}, we plot the relative changes of the $220$, $330$, and $210$ mode amplitudes as functions of the reference eccentricity $e$. It can be seen that for the considered eccentricity range, the increase of the eccentricity tends to give a smaller excitation amplitude, but the trend is weak. We also note that the apparent change might be caused by the extraction uncertainty. Nevertheless, the largest change in the amplitudes should be less than $5\%$.

\begin{figure}[htbp]
  \centering
  \includegraphics[width=0.45\textwidth]{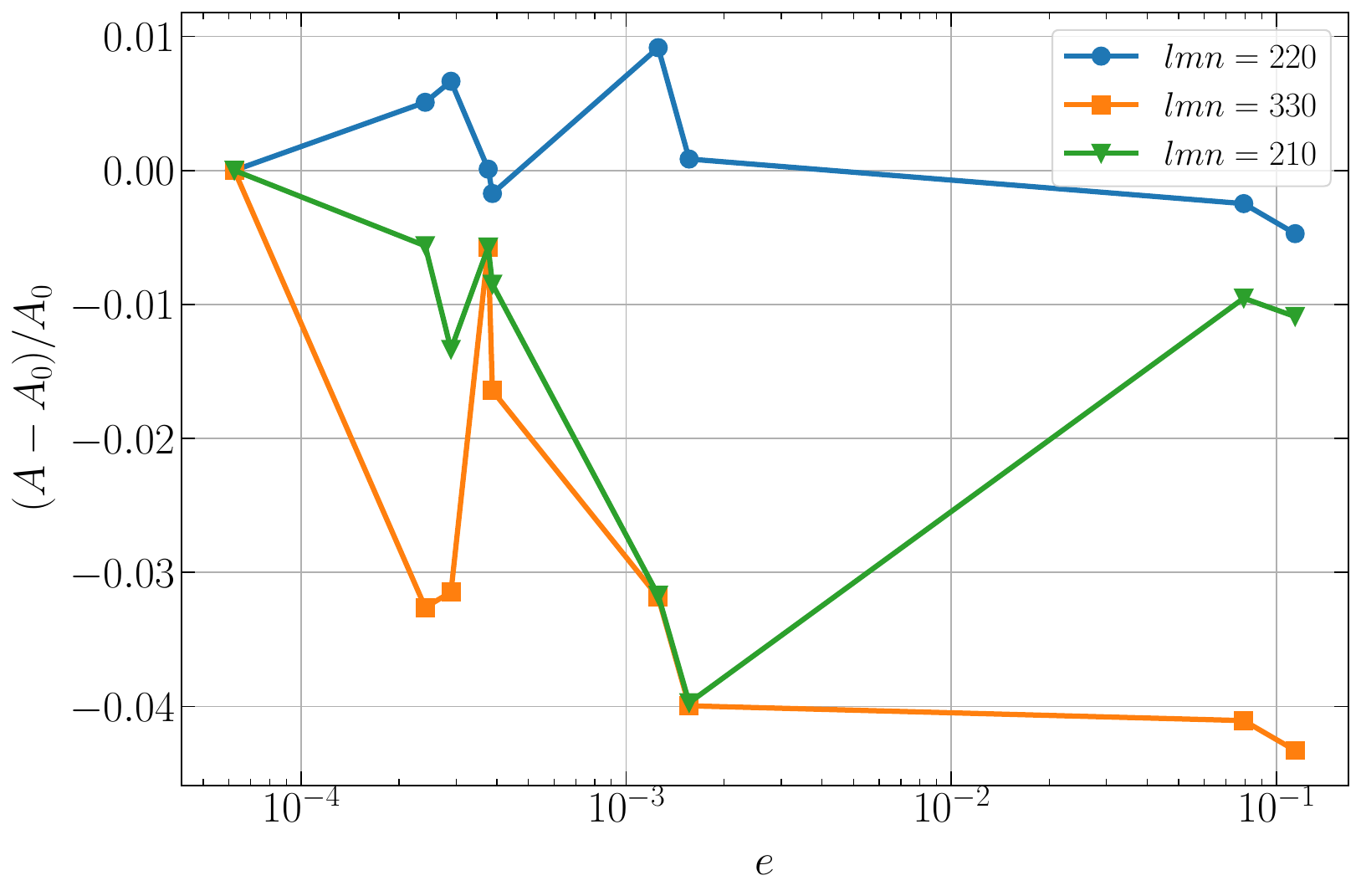}
  \caption{The relative changes of the $220$, $330$, and $210$ mode amplitudes $(A-A_0)/A_0$ as functions of the reference eccentricity $e$. $A_0$ is the mode amplitude of the simulation with the lowest eccentricity.  \label{fig:sxs_q2}}
\end{figure}

According to the results shown in the following sections, the above estimation suggests that the effect of eccentricity does not affect our conclusions, even if we do not apply the eccentricity reduction procedure for our simulations. Therefore, considering the computational cost, in the following discussion, all the simulations with different $\lambda$ adopt the same GR initial data. To further confirm this point, in Appendix~\ref{app:ecc}, we also perform eccentricity reduction to the simulation with the largest eccentricity (also the largest $\lambda$) and analyze the effect of eccentricity. We note that, even with the GR initial data that suffer from an initial growth of the scalar field, we are still able to reduce the eccentricity with the procedure discussed in the Appendix.

\subsection{Simulations and QNM extraction results}

We perform six simulations with coupling constant $\lambda/M^2=0$, $0.01$, $0.02$, $0.03$, $0.04$, and $0.05$. The largest coupling normalized to the mass of small BH $\lambda/m_2^2=0.2$ is close to the limit given by the loss of hyperbolicity. Simulation with a higher coupling $\lambda/M^2=0.06$ can be performed without a crash of the code. However, we find strong constraint violations outside the AH, so we discard this simulation. On the other hand, we find that $\lambda/M^2=0.07$ seems to be larger than the allowed coupling. Thus, the evolution inevitably leads to a non-hyperbolic region outside the AH and our code crashes. 

\begin{figure}[htbp]
  \centering
  \includegraphics[width=0.45\textwidth]{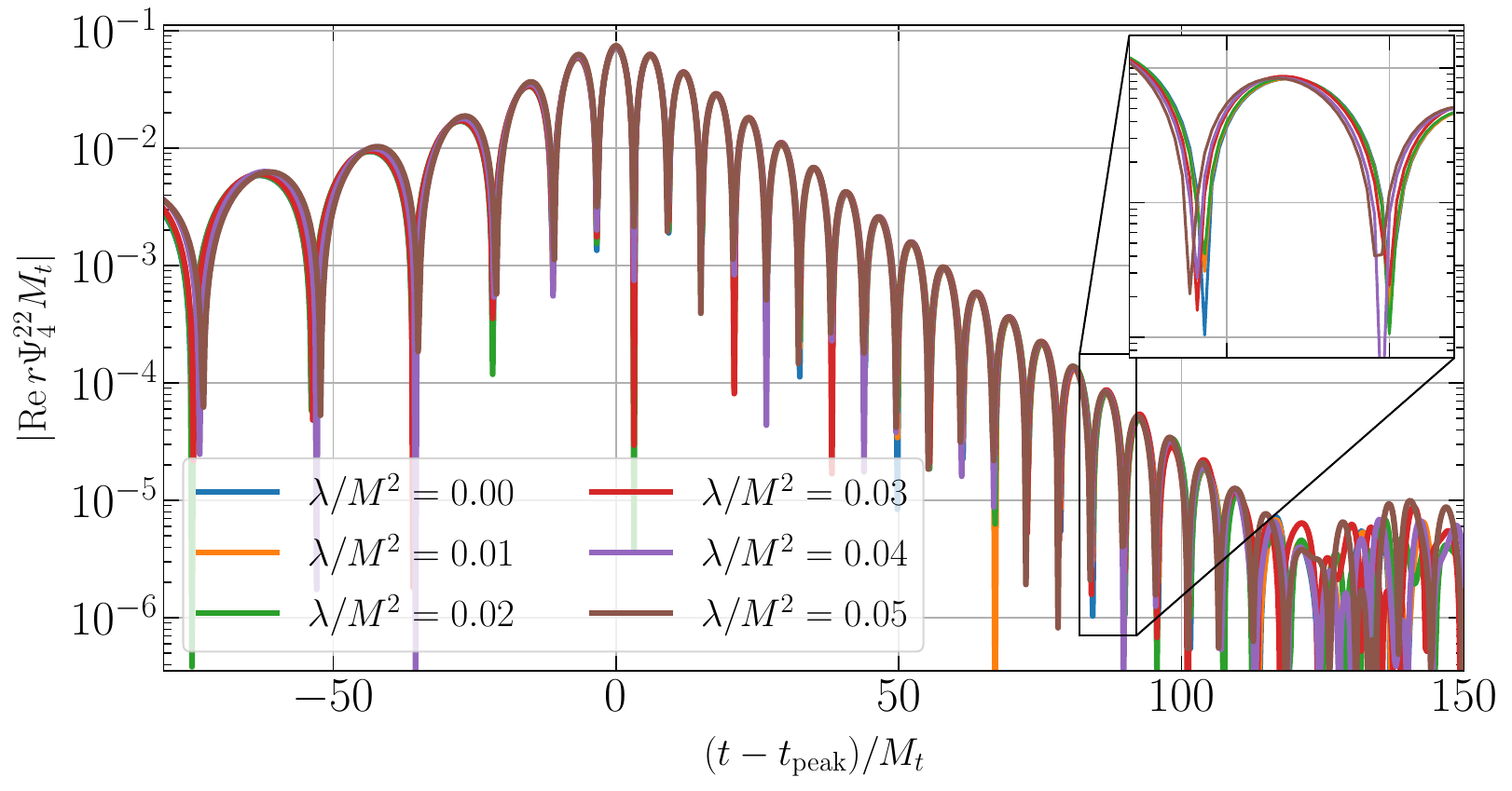}
  \caption{The real part of $\Psi_4^{22}$ waveforms for the simulations with different $\lambda$. We align them at the peak of $|\Psi_4^{22}|$ and adjust the phase to be zero at that time. The upper right small figure is a zoom in of part of the ringdown. \label{fig:psi4_22}}
\end{figure}

In Fig.~\ref{fig:psi4_22}, we show the real part of $\Psi_4^{22}$ waveforms of the six simulations. They have been aligned at the peak of $\Psi_4^{22}$, which means that we adjust the phases of the waveforms to be zero at their peak time. The $lm=22$ waveform only shows a very small difference, even for the largest coupling case. From the zoom-in figure in the upper right part, we can see that the scalar coupling leads to an increase in the real part of the mode frequency, which, according to later discussion, is mainly caused by the change of the final mass of the remnant BH.

\begin{figure}[htbp]
  \centering
  \includegraphics[width=0.45\textwidth]{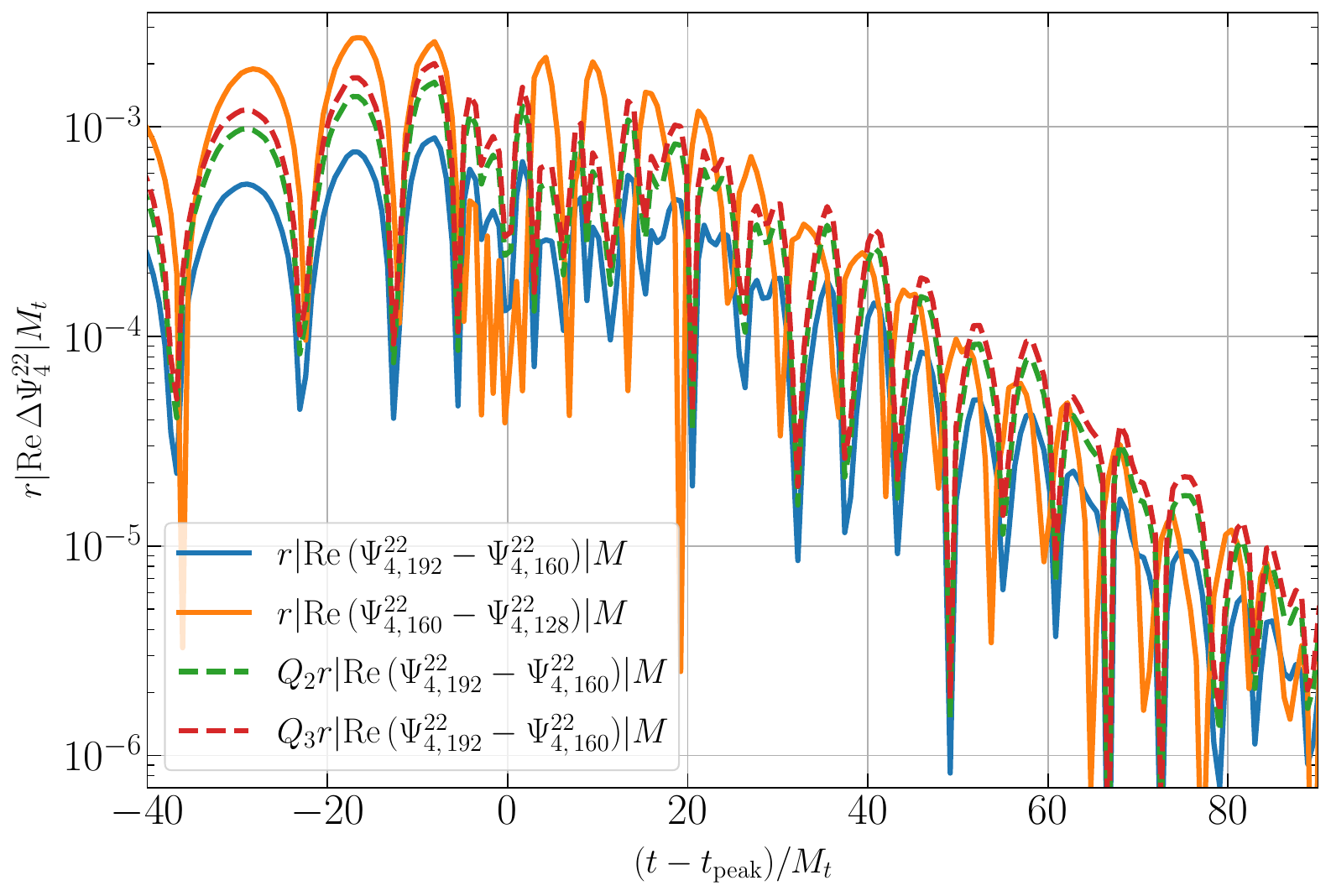}
  \caption{Convergence of the waveform ${\rm Re}\,\Psi_4^{22}$ for BBH merger simulation with coupling $\lambda/M^2=0.05$. The simulations have finest grid resolutions $\Delta x=M/64$, $M/80$, and $M/96$ and are marked by subscripts $128$, $160$, and $192$ respectively. The multiplying coefficient corresponding to a second-order convergence is $Q_2=(1/128^2-1/160^2)/(1/160^2-1/192^2)$. The figure shows a second to third-order convergence for the waveform $\Psi_4^{22}$ around the peak time. We only show a short time window in order for clarity. \label{fig:BBH_convergence}}
\end{figure}

Before we proceed, we first present the convergence test for our BBH merger simulations. We perform the test for the simulation with the largest coupling $\lambda/M^2=0.05$. We use the finest grid resolution of $\Delta x=M/64$, $M/80$, $M/96$ and they are denoted by the subscripts $128$, $160$, and $192$, respectively. In Fig.~\ref{fig:BBH_convergence}, we present the convergence of the waveform ${\rm Re}\,\Psi_4^{22}$ around the merger and during the ringdown. We aligned the three simulations at the peak of $|\Psi_4^{22}|$. It can be seen that the waveform shows a second to third-order convergence, which is consistent with the convergence order for \texttt{GRChombo}~\cite{Radia:2021smk}.

To extract the QNM amplitudes and phases, we apply the extraction procedure introduced in Sec.~\ref{sec:extract procedure} to our simulations. We find that for frequency-agnostic fitting, for a given $lm$, only one mode appears with frequency close to the estimated frequency of the $lm0$ mode. That is, no mode splitting is observed, though the polar and axial modes are expected to have different frequencies. The possible main reason is that we applied the reflection symmetry in the simulation, which suppresses the existence of the axial modes~\cite{Martel:2005ir}. Therefore, in our extraction results, all modes are assumed to be polar modes. However, as we discussed in Sec.~\ref{sec:single BH}, even if there is excitation of polar and axial modes simultaneously in some systems, for example, systems with precession, due to the very close frequencies of the two modes, it is challenging to extract them separately. 

\begin{figure}[htbp]
  \centering
  \includegraphics[width=0.45\textwidth]{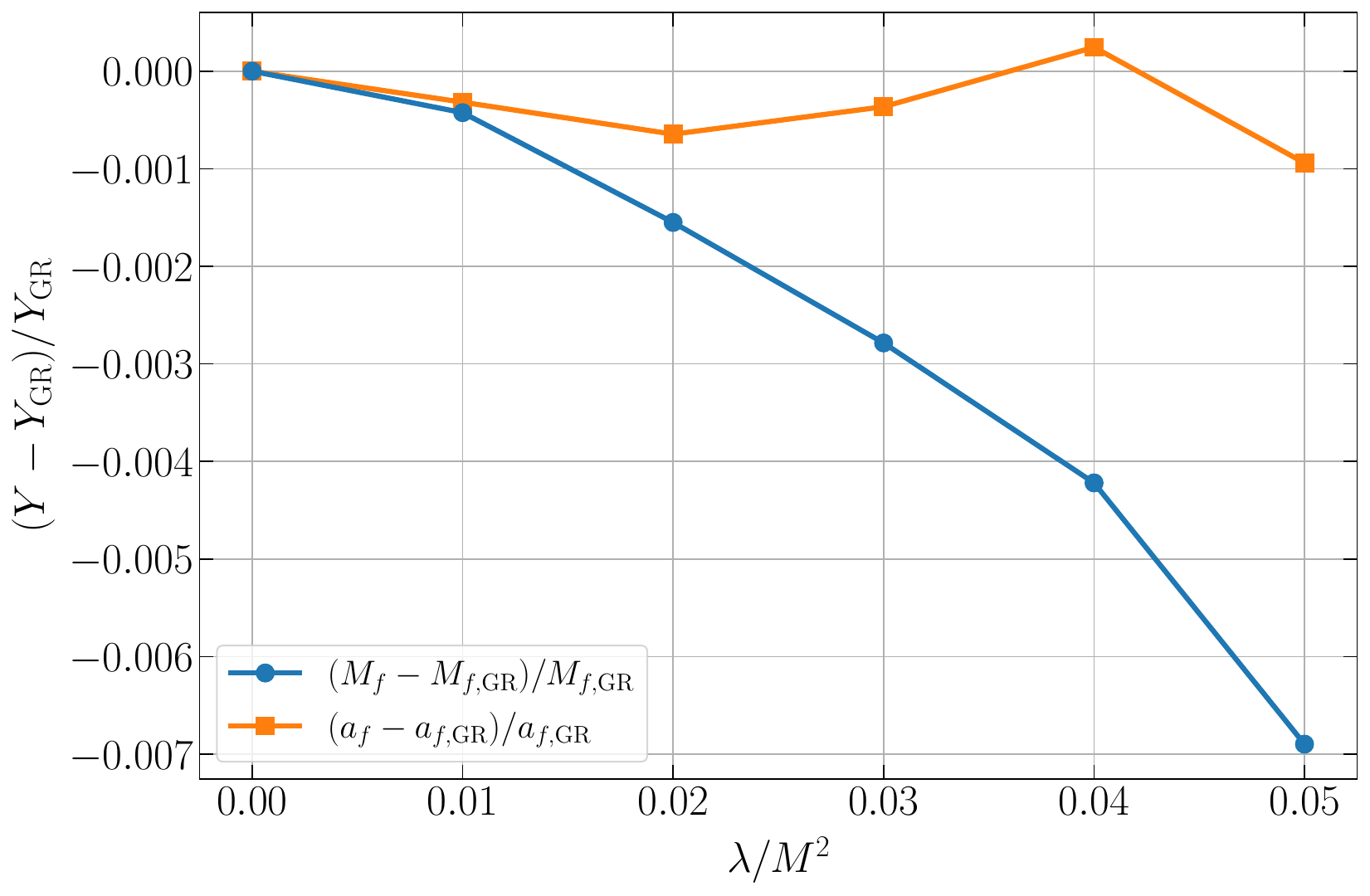}
  \caption{We calibrate the mass and spin of the remnant with the frequency fitting results from the $220$ mode. $Y$ in the y-axis label stands for the remnant mass or spin, while $Y_{\rm GR}$ is the value of the $\lambda=0$ case. \label{fig:MS_from_220}}
\end{figure}

Fitting a mode with a fixed mode frequency generally provides a more stable result of the mode amplitude and phase compared to a frequency-agnostic fitting. Therefore, to take advantage of the theoretically calculated mode frequencies~\cite{Chung:2024vaf}, we need to obtain the ADM mass and spin of the remnant BH, which can be different from the mass and spin extracted at the AH using the Christodoulou formula~\cite{Christodoulou:1970wf,Christodoulou:1971pcn} due to the presence of a scalar field. Instead of integrating an additional code for extracting ADM quantities, we find that it is convenient and accurate enough to do it in the following way: we calibrate the values of the mass and the spin of the remnant BH with the $220$ mode frequency. As we know the coupling strength $\lambda/M^2$ and assume that there are only polar modes, the mode frequency then is only a function of the unknown ADM mass and spin of the remnant BH, $\omega^p(M_f,a_f,\lambda)$, where the superscript ``$p$'' denotes the polar mode. A fitting formula for $\omega^p$ is given in Ref.~\cite{Chung:2024vaf}. We then perform a frequency-agnostic fit for the $220$ mode. According to our experience in GR simulations, the frequency fitting of the $220$ mode can, in general, reach a precision of $0.1\%$ in both real and imaginary parts. Therefore, using the fitting result of the $220$ mode frequency, we can solve for the mass and spin of the remnant BH, which provides an accurate enough result. In Fig.~\ref{fig:MS_from_220}, we plot the relative change of the remnant mass and spin as functions of the coupling $\lambda$. The remnant mass shows a clear trend of decrease as the coupling $\lambda$ increases, caused by the increasing scalar radiation, while the remnant spin shows no clear trend and its variation is within the extraction precision. The decrease of the remnant mass is also suggested by Fig.~\ref{fig:psi4_22}, where one can see that the waveforms of larger $\lambda$ have higher real frequencies during the ringdown, even though the polar mode tends to give a smaller real frequency when $\lambda$ increases (for example, see Fig.~\ref{fig:P_33}). 

\begin{figure}[htbp]
  \centering
  \includegraphics[width=0.45\textwidth]{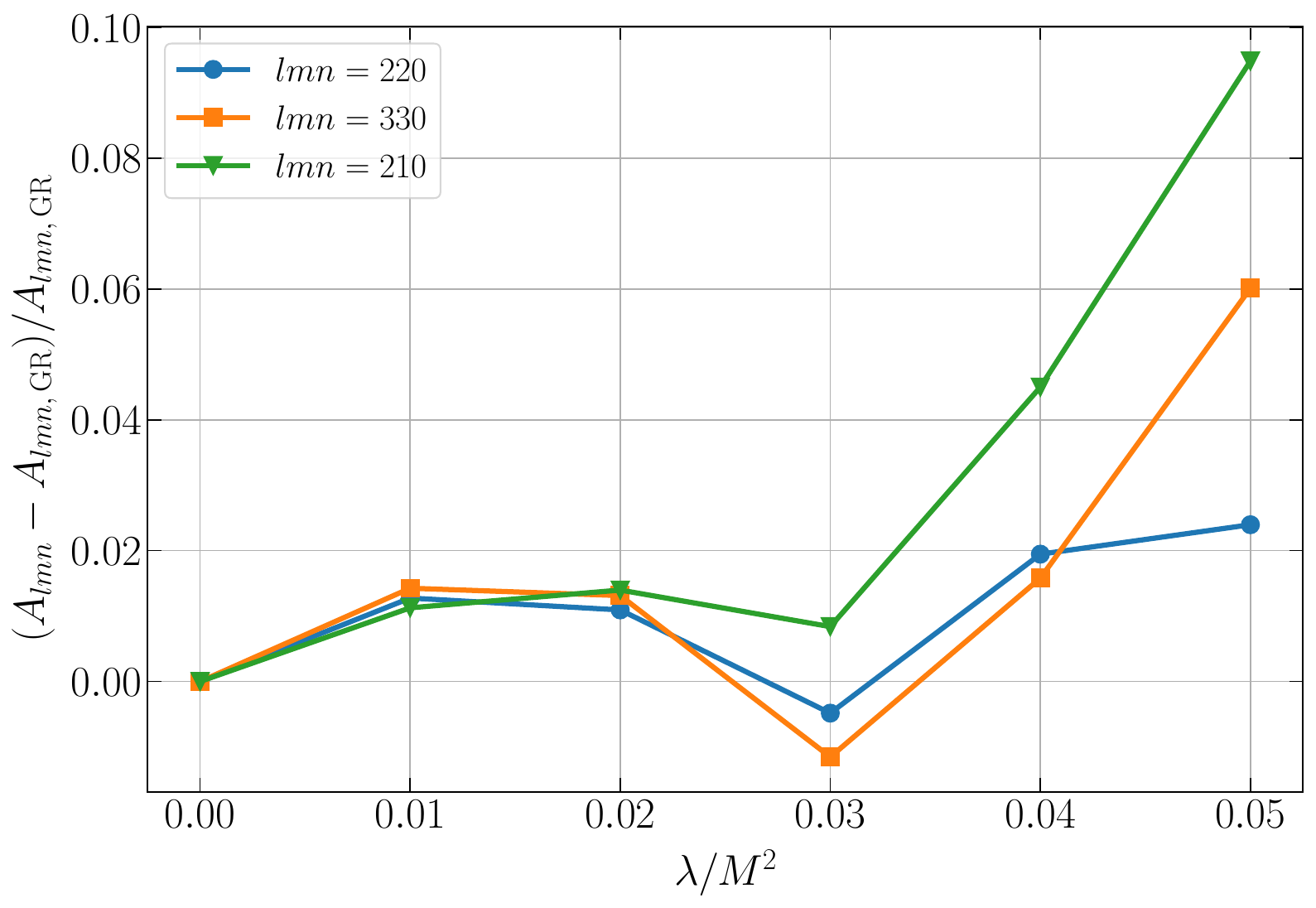}
  \caption{The relative change of the mode amplitudes as functions of $\lambda$ for $220$, $330$, and $210$ modes. For the fitting, we fix the mode frequency for the one desired mode and add additional free modes with $N_{\rm free}=5$.  \label{fig:A_fix_3modes}}
\end{figure}

\begin{figure}[htbp]
  \centering
  \includegraphics[width=0.45\textwidth]{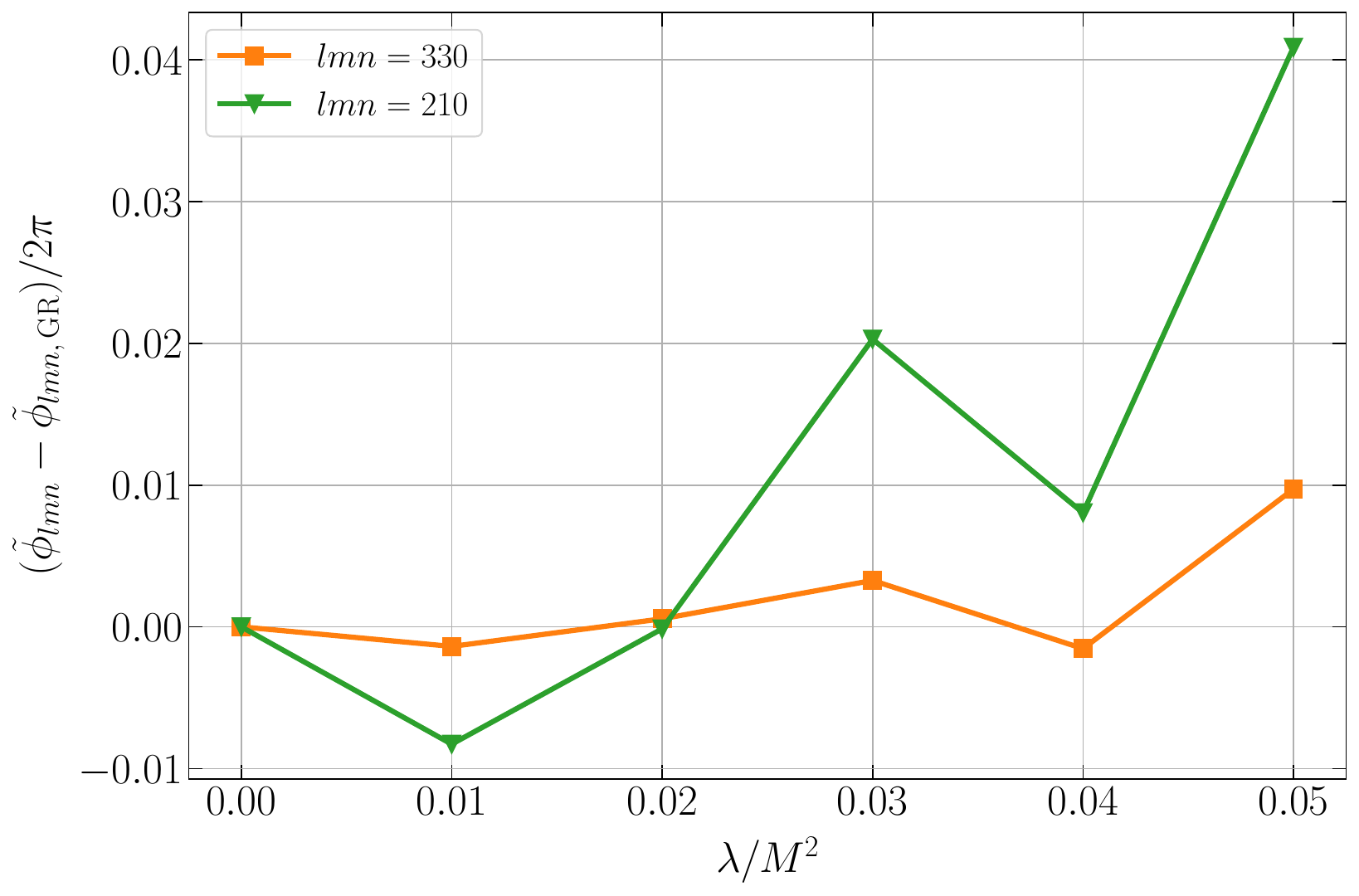}
  \caption{The change in the adjusted phase $\tilde{\phi}_{lmn}=2\phi_{lmn}-m\phi_{220}$ for $330$, and $210$ modes.  \label{fig:phi_fix_2modes}}
\end{figure}

With the calibrated mass and spin of the remnant BH and the theoretically calculated mode frequencies~\cite{Chung:2024vaf}, we perform fixed frequency fitting for the $220$, $330$, and $210$ modes. In the fitting, we also add additional free modes to stabilize the desired mode and we use $N_{\rm free}=5$. Figure~\ref{fig:A_fix_3modes} shows the relative changes in the mode amplitudes as functions of the coupling $\lambda$, while Fig.~\ref{fig:phi_fix_2modes} shows the change in the adjusted phase $\tilde{\phi}_{lmn}=2\phi_{lmn}-m\phi_{220}$~\cite{Cheung:2023vki}. From Fig.~\ref{fig:A_fix_3modes} one can observe a weak trend that the mode amplitudes increase as the coupling $\lambda$ increases. However, considering that the largest coupling we used is already at the border of the loss of hyperbolicity and out of the weak coupling regime where the sGB gravity can be regarded as an EFT, the change in the mode excitation is rather small. Even for the largest coupling, the change of the mode amplitudes is around $10\%$ and the change of the adjusted phase is less than $0.3\,{\rm rad}$. Further, the change in the dominant $220$ mode amplitude is only up to around $2\%$. We also note that, according to the discussion in Sec.~\ref{subsec:ecc} and Appendix~\ref{app:ecc}, the small eccentricity in our simulations does not alter our conclusion here. 

We should mention again that we are using improper initial data. Starting from the GR initial data leads to a violent scalarization process, introducing an additional orbital eccentricity. Further, this scalarization process also changes the ADM mass of the BHs. We expect that the change of mass caused by scalarization will not be larger than the mass change shown in Fig.~\ref{fig:MS_from_220}, but it still leads to an improper comparison. Therefore, studies of constructing initial data that are consistent with the field equations are important for future simulations~\cite{Nee:2024bur}.

\begin{figure}[htbp]
  \centering
  \includegraphics[width=0.45\textwidth]{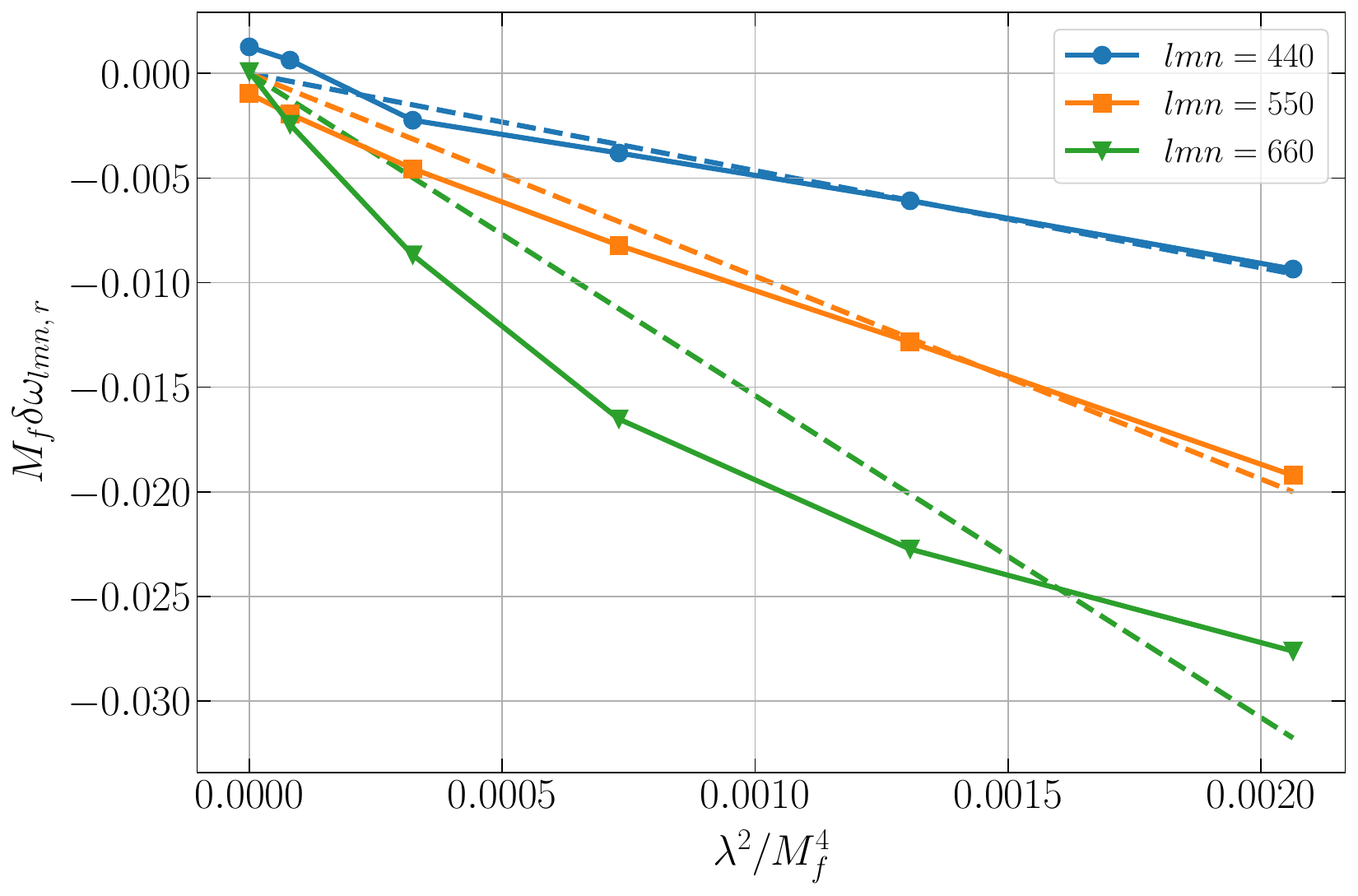}
  \caption{The change of the real part of the $440$, $550$, and $660$ mode frequencies, $\delta\omega_{lmn,\,r}=\omega_{lmn,\,r}-\omega_{lmn,\,r,\,{\rm GR}}$, as functions of the coupling $\lambda$. Note that we normalize $\lambda$ and $\delta\omega_{lmn,\,r}$ with $M_f$, as for simulations with different $\lambda$, they have slightly different remnant masses. The dashed lines show the linear fitting of $\delta\omega_{lmn,\,r}$. For the extraction, we use $N_{\rm free}=5$. \label{fig:omega_r_free_3modes}}
\end{figure}

For modes other than $220$, $330$, and $210$, we can only perform frequency-agnostic fittings since numerical values are not available in the literature \cite{Chung:2024vaf}. In Fig.~\ref{fig:omega_r_free_3modes}, we show the difference between the mode frequencies obtained from the fitting and the GR predictions for the $440$, $550$, and $660$ modes. At the leading order, the deviation of the mode frequencies from their GR counterpart is proportional to $\lambda^2$, that is, $M_f\delta\omega=k(a_f/M_f)\lambda^2/M_f^4$~\cite{Chung:2024vaf}, where $k(a_f/M_f)$ is a coefficient that depends on the BH spin. In our fitting result, we expect that the change of mode frequency is roughly proportional to $\lambda^2$ as the final spin of the remnant BH in our simulations only changes slightly for different couplings as shown in Fig.~\ref{fig:MS_from_220}. It is clear that for the $440$ and $550$ modes, the change of the real part of the mode frequencies roughly follows a straight line as shown by the dashed lines in Fig.~\ref{fig:omega_r_free_3modes}. The mode $660$ shows a different result, probably due to the extraction uncertainties. Higher modes are generally harder to extract because of their smaller excitation. Also, in principle, the imaginary part of the mode frequency should follow the same discussion as above. In our extraction procedure, we have roughly the same absolute precision for the real and imaginary parts of the mode frequency. However, the imaginary part of the mode frequency is in general one order of magnitude smaller than the real part for $n=0$ modes. The extraction precision for those modes shown in Fig.~\ref{fig:omega_r_free_3modes} seems to be insufficient to capture the change in the imaginary part of the mode frequencies. 

\begin{figure}[htbp]
  \centering
  \includegraphics[width=0.45\textwidth]{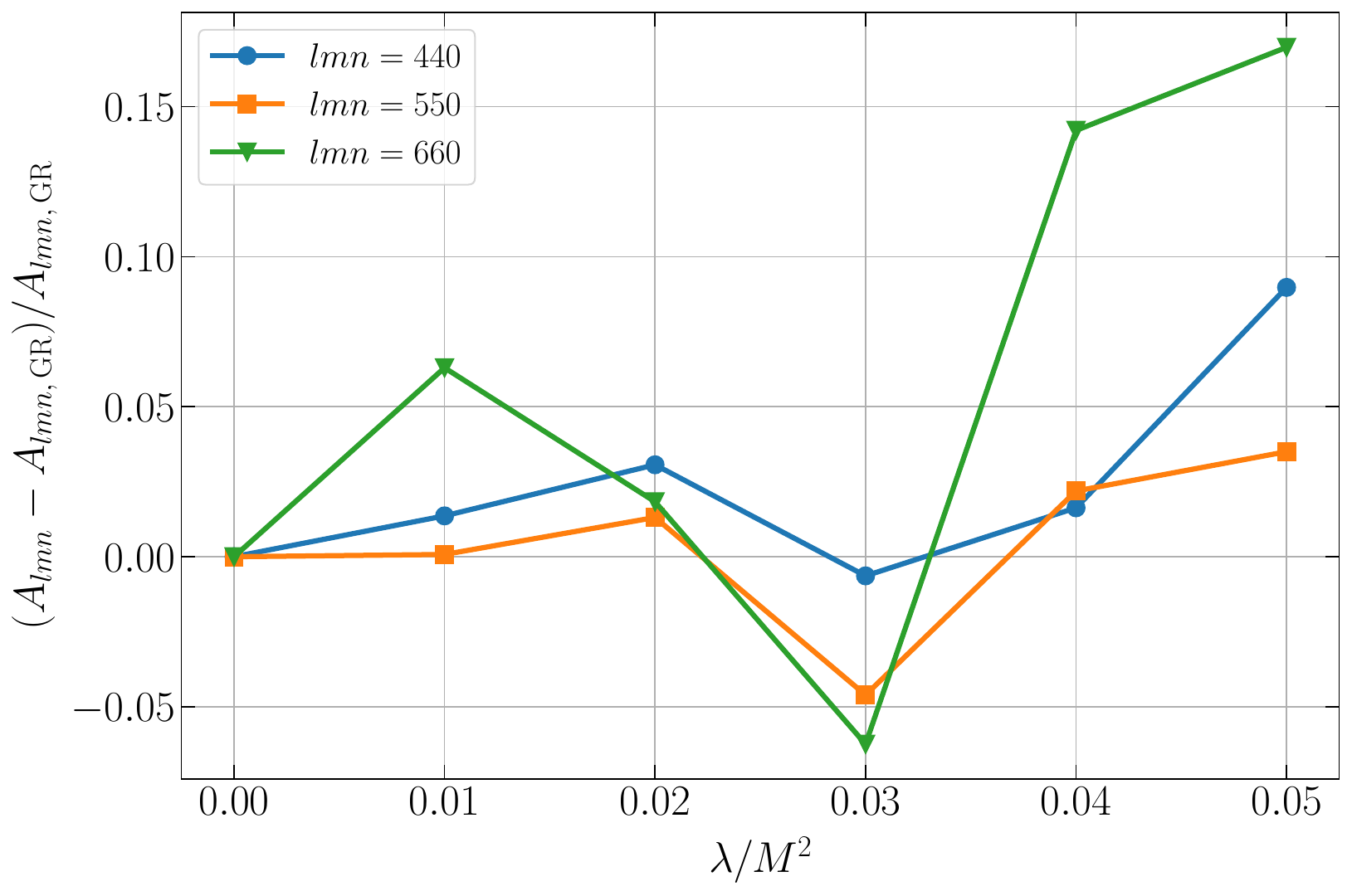}
  \caption{Similar as Fig.~\ref{fig:A_fix_3modes} but for the $440$, $550$, and $660$ modes. Note that the modes shown in this figure are fitted in the frequency-agnostic way.  \label{fig:A_free_3modes}}
\end{figure}

Finally, we show the change of the mode amplitudes for the $440$, $550$, and $660$ modes in Fig.~\ref{fig:A_free_3modes}. It shows a similar trend as in Fig.~\ref{fig:A_fix_3modes}, though these modes are fitted in a frequency-agnostic way. The largest change is again around $10\%$, while for the $660$ mode it seems to have a change larger than $15\%$. Still, one should keep in mind that the $660$ mode also has the largest extraction uncertainty.

The excitation of higher overtones ($n\geq 1$) can also be an interesting topic as they have been shown to be important in ringdown analysis~\cite{Giesler:2019uxc}. However, they have a short damping time and are hard to extract with frequency-agnostic fitting. Therefore, we do not show the extraction result here. One could expect that their excitation will also not change much as in those modes we have shown before. Nevertheless, calculations of the mode frequencies of these overtones based on the solution of the linearized perturbation equations are important for a detailed study of their excitation from numerical simulations.

As mentioned above, recent studies also showed that, for robust tests of gravity in the ringdown regime, it is necessary to accurately model the extra field-induced modes~\cite{Crescimbeni:2024sam,Crescimbeni:2025kxi}, like the scalar-led modes in sGB theory~\cite{Blazquez-Salcedo:2024oek}. These modes have excitation amplitudes at the order of $O(\lambda^2/M^4)$, which are in principle at the same order as the change of the amplitudes of the tensorial modes. Therefore, we also perform extraction for the scalar-led $220$ mode. We employed both frequency-agnostic fitting and fixed-frequency fitting. For the frequency-agnostic fitting, as one may expect, due to the very small amplitude of the scalar-led mode, there is no clear evidence around the expected frequency. For fixed-frequency fitting, however, we can only use the mode frequency at the test-field limit and for a Kerr spacetime. As mentioned in the previous section, the change in the mode frequencies of the scalar-led modes is at the same order as the tensor modes. Using an incorrect mode frequency makes the extraction more unstable, and we do not obtain trustable results for the scalar-mode amplitudes.

\section{Spin-induced and dynamical scalarization}\label{sec:phi2}
In previous sections, we mainly focused on the shift-symmetric sGB theory, in which a static BH always carries a scalar charge~\cite{Sotiriou:2014pfa}. Although we have shown that the ringdown excitation of BBH merger in this theory is not strongly affected by the scalar coupling, the whole waveform, including the inspiral part, can still be largely affected due to the additional scalar radiation~\cite{East:2020hgw,AresteSalo:2025sxc} and therefore can be well constrained through full inspiral-merger-ringdown analysis. However, in the sGB gravity with quadratic coupling, there can exist phenomena called spin-induced scalarization and dynamical scalarization~\cite{Doneva:2017bvd,Silva:2017uqg,Antoniou:2017acq,Elley:2022ept,Doneva:2023oww}. In the former case, scalarization is a result of the change of the Gauss-Bonnet invariant sign around the poles that happens for sufficiently rapidly rotating BHs. For dynamical scalarization, the scalar field development is a result of the collective effect of the two BHs in a binary. Namely,  in some region of the parameter space, the BH will not scalarize when insulated. If binary separation decreases enough, though, scalar field development might be triggered. In this section, we start with a coupling that leads to the spontaneous spin-induced scalarization and also discuss the case of dynamical scalarization at the end.

For the case of spin-induced scalarization, as we discussed, a BH only scalarizes when it has a large spin; otherwise, it adopts the GR solution. In such a theory, BBH mergers with zero or low progenitor spins can be indistinguishable from GR as the two initial BHs do not carry scalar charge. It only deviates from GR in the post-merger stage when the fast-spinning remnant triggers the scalarization. Such a scenario can only be tested through ringdown observations. In this section, we investigate the ringdown in such a theory and give an upper limit of the deviations of the ringdown waveform from GR.

We perform BBH merger simulation in sGB gravity with a coupling function shown by Eq.~(\ref{eq:cop_phi2}) with $\epsilon=-1$, namely
\begin{equation}
    f(\varphi)=-\frac{\lambda}{2\beta}\left(1-e^{-\beta\varphi^2}\right)\,.
\end{equation}
The particular sign of $\epsilon$ is chosen such that it allows for spin-induced scalarization. 
In this function, the constant $\lambda$, which is effectively the coefficient of the $\varphi^2$ term if one expands the exponential term in the limit of small scalar field, mainly controls the onset of the scalarization and the growth rate of the scalar field as the effective mass of the scalar field is proportional to $\lambda$~\cite{Dima:2020yac}
\begin{equation}
    \mu_{\rm eff}=\lambda\mathcal{R}_{\rm GB}^2\,.
\end{equation}
When $\mu_{\rm eff}<0$, the scalar field suffers from the tachyonic instability and the GR solution is no longer favored. For a Schwarzschild BH, the Gauss-Bonnet invariant is positive everywhere, but for a spinning BH, $\mathcal{R}_{\rm GB}^2$ can be negative around the pole. Therefore, for $\lambda>0$, only spinning BHs can trigger the instability and obtain a scalar charge. A greater $\lambda$ leads to a lower BH spin threshold for scalarization and a shorter instability time scale~\cite{Dima:2020yac}.

The constant $\beta$ is introduced as the coefficient related to $\varphi^4$ and higher-order terms. Only quadratic coupling leads to an unstable GR BH solution, but higher-order self-interaction of the scalar field can stabilize the scalarized solutions~\cite{Blazquez-Salcedo:2024oek,Silva:2018qhn}. This nonlinear term determines the end state of the scalarization process. For spin-induced scalarization, a larger $\beta$ suppresses the scalarization and leads to a smaller scalar field in the end state~\cite{Doneva:2023oww}.

In a specific range of parameters, the BBH merger simulation we studied with mass ratio $q=1.2$ and total initial mass $M_t=1.1\,M$, can exhibit the spin-induced dynamical scalarization. As the progenitor BHs are non-spinning, we only need to adjust the parameter so that the remnant BH can be scalarized. In our simulations, the final spin of the remnant BH is around $a_f/M_f\sim 0.68$. According to linear analysis, one needs to have $\lambda/M_f^2\gtrsim 10$ to trigger the instability~\cite{Dima:2020yac}. Therefore, we choose $\lambda/M^2=20$ and $\lambda/M^2=50$ as two representative examples that we will discuss in detail later. 

For a given $\lambda$, as we mentioned above, $\beta$ controls the final amplitude of the scalar field. Similar to the shift-symmetric case, a too large scalar field that exits the weak-coupling regime will lead to the loss of hyperbolicity~\cite{Doneva:2023oww}. For each $\lambda$, there exists a minimum $\beta$ that gives the largest scalar field while keeping the non-hyperbolic region inside the AH. As we aim to study the possible largest deviation in the ringdown stage, we fine-tune the parameter $\beta$ as well as the cutoff parameters introduced in Eq.~(\ref{eq:cutoff}) to find the minimum value of $\beta$ (thus the maximum scalar field) that is allowed in our simulations.

\begin{figure}[htbp]
  \centering
  \includegraphics[width=0.45\textwidth]{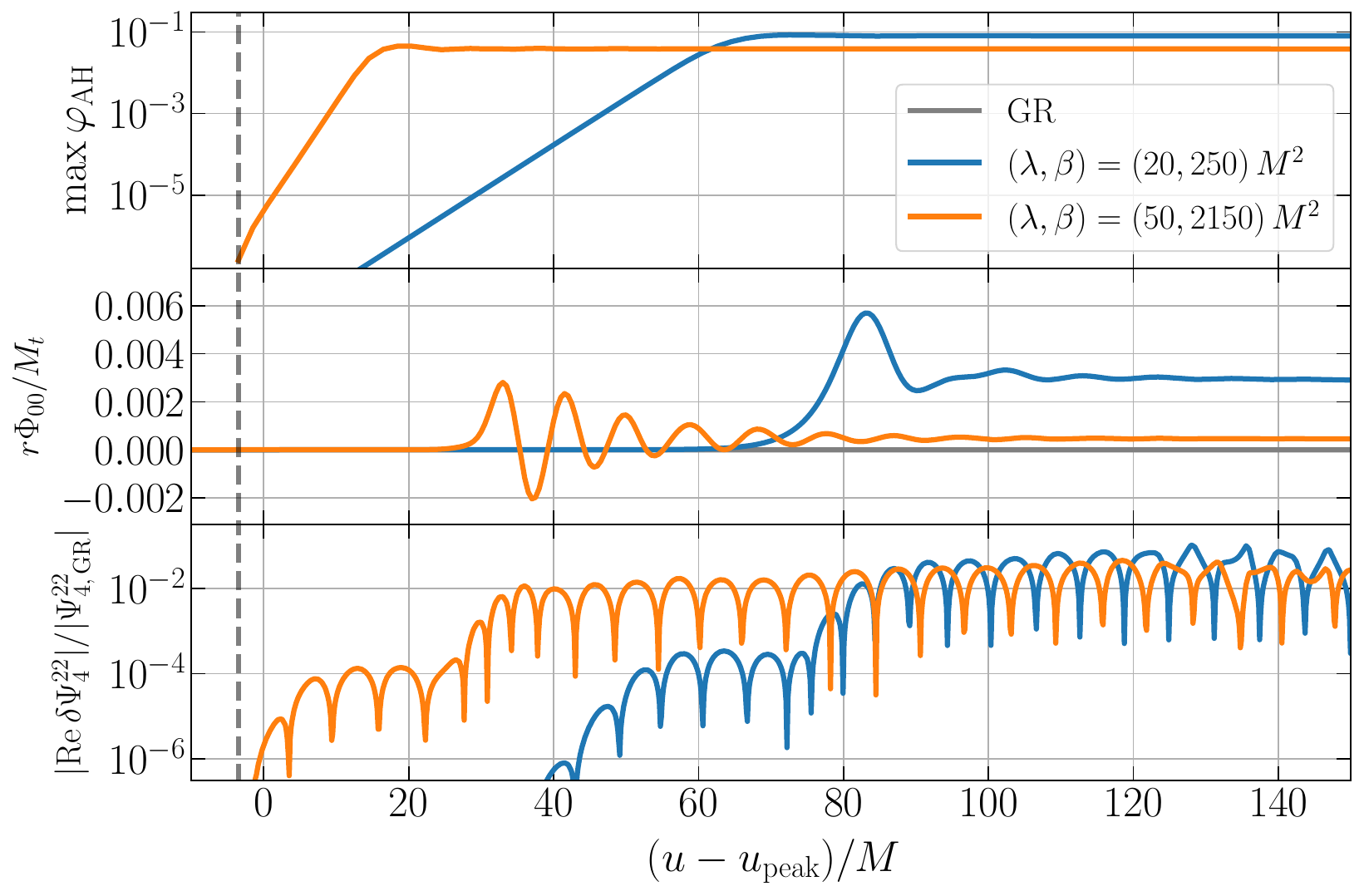}
  \caption{Evolution of the maximum scalar field at the AH, scalar field monopole $\Phi_{00}$, and the relative change of $\Psi_4^{22}$ caused by the spin-induced scalarization. $\delta\Psi_{4}^{22}=\Psi_4^{22}-\Psi_{4,\,{\rm GR}}^{22}$, $u=t-r$ is the Bondi time, and $u_{\rm peak}$ denotes the peak time of $|\Psi_{4,\,{\rm GR}}^{22}|$. The vertical dashed line denotes the formation of the common AH. It is clear that a deviation from GR in $\Psi_4^{22}$ starts developing when the scalarization is triggered.  \label{fig:spin_induce}}
\end{figure}

In Fig.~\ref{fig:spin_induce}, we show the evolution of the maximum scalar field at the AH, scalar monopole $\Phi_{00}$, and relative change of $\Psi_4^{22}$ caused by the spin-induced scalarization at the ringdown stage. Note that for spin-induced scalarization, the maximum of the scalar field is located at the pole of the BH~\cite{Doneva:2023oww}. We use the Bondi time $u=t-r$ so that one can roughly compare $\Phi_{00}$ and $\Psi_4^{22}$, that are extracted at $r=90\,M$, with $\varphi_{\rm AH}$, extracted at the AH. $u_{\rm peak}$ is the peak time of the $|\Psi_{4,\,{\rm GR}}^{22}|$ and roughly around the time of merger. One can clearly see that, after the merger, there is a growth of the scalar field at the AH. For $\lambda/M^2=50$, the growth time of the scalar field is around $20\,M$, while for $\lambda/M^2=20$, the growth of the scalar field takes around $70\,M$. We have adjusted $\beta$ to have a maximum scalar field for both $\lambda$ without losing hyperbolicity. The final scalar field for $\lambda/M^2=20$ is around twice as large as in the $\lambda/M^2=50$ case. The GW waveform $\Psi_{4}^{22}$ is affected by the growth of the scalar field and deviates from GR when the scalar field grows. 

Although we have fine-tuned the parameter to have a maximum scalar field, the maximum relative deviation of $\Psi_{4}^{22}$ from GR is only around $1\%$ for the main part of the ringdown. The final relative deviation for $\lambda/M^2=20$ is larger due to the larger scalar field in this case. However, the long growth time of the scalar field for a small coupling $\lambda$ will cause the deviation from GR to become hard to observe as the ringdown decays exponentially. This $1\%$ level deviation is similar to the shift-symmetric case, where the dominant $220$ mode amplitude only changes around $2\%$ for the largest coupling we considered. 
Thus, for the spin-induced scalarization scenario, the QNM amplitudes in the ringdown part differ from GR by a small but non-negligible amount. Though the mode amplitude may change only slightly, BH spectroscopy might still constrain the theory, as the frequency measurements are better than the amplitude measurements in general. Therefore, it will be interesting to study the effects of QNM frequency evolution in the ringdown part as there is a growing scalar field after the merger. However, this is out of the scope of this paper and we leave it for future study.

Different from the spin-induced scalarization discussed above, for the coupling in Eq.~(\ref{eq:cop_phi2}) with $\epsilon=+1$ and positive $\lambda>0$, there exists the normal curvature-induced scalarization scenario, as for a non-rotating BH the Gauss-Bonnet invariant is positive everywhere~\cite{Doneva:2017bvd,Silva:2017uqg,Antoniou:2017acq}. Recent studies \cite{Julie:2023ncq,Capuano:inprep} found that, for a particular region of the parameter space, there can exist a dynamical scalarization process during BBH merger similar to the neutron star binary dynamical scalarizaton \cite{Barausse:2012da,Shibata:2013pra}. As we discussed above, when the two BHs are far from each other, they are both not scalarized. However, when they approach each other, a scalar field develops. After the merger, as the dimensionless coupling $\lambda/M_f^2$ for the remnant BH is much smaller, the spacetime will be a Kerr spacetime at the end. In such a scenario, the QNM frequencies will be the same as in GR. Therefore, BH spectroscopy will be blind to this parameter space of the sGB theory. Nevertheless, one may expect that due to the scalar field developed in the late inspiral and merger, the QNM amplitude and phase will still be different.

\begin{figure}[htbp]
  \centering
  \includegraphics[width=0.45\textwidth]{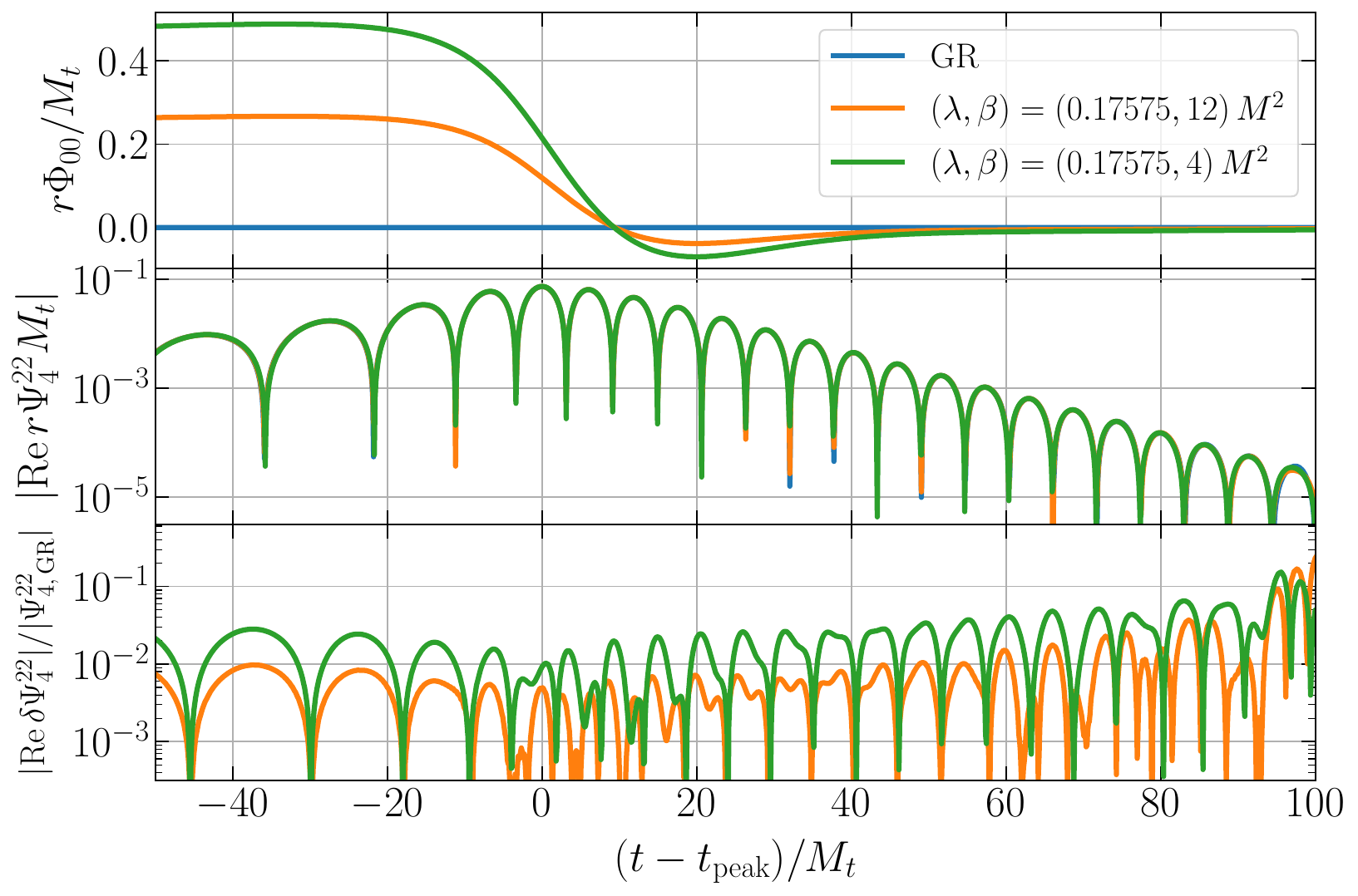}
  \caption{Evolution of the scalar field monopole $\Phi_{00}$, waveform $\Psi_4^{22}$ and the relative change of $\Psi_4^{22}$ as functions of the time $t$. We align the waveform around the peak. The scalar field is dynamically excited and descalarized after merger. \label{fig:dynamical}}
\end{figure}

Therefore, we also perform simulations for BBHs in the sGB theory within this specific parameter space. The parameter space of $\lambda$ that allows for dynamical scalarization process is quite limited since the value of $\lambda/M^2$ should be close to (but smaller than) the threshold one for which isolated BHs develop scalar field. According to the detailed studies in Ref.~\cite{Capuano:inprep}, we fix $\lambda/M^2=0.17575$ and consider an equal-mass, non-spinning BBH so that $q=1$ and $M_t=1\,M$. Similar to the spin-induced scalarization case, the parameter $\beta$ controls the maximum scalar field, and we choose two values, $\beta=12$ and $\beta=4$, where the latter one is close to the limit of loss of hyperbolicity. In Fig.~\ref{fig:dynamical}, we show the evolution of the scalar field monopole $\Phi_{00}$, waveform $\Psi_4^{22}$ and the relative change of $\Psi_4^{22}$ as functions of the time $t$. One can see that the scalar field quickly drops to zero after the merger. Though the late inspiral and merger are affected by the development of the scalar field (for detailed discussions, see Ref.~\cite{Capuano:inprep}), after we align the waveform around the peak, the difference caused by the scalar field is only around $2$ to $3$ percent level for the main part of the ringdown, which is similar to the change of the dominant $220$ mode amplitude in the shift-symmetric case. The fact that the remnant is Kerr-like and thus the QNM frequencies are the same, while the ringdown mode amplitudes change slightly, may cause difficulties in using ringdown observation to constrain dynamical scalarization through future ringdown GW observations.

\section{Conclusions}\label{sec:conclusions}

In this paper, we study the ringdown of BBH mergers in sGB gravity. By performing $3+1$ NR simulations of single perturbed BHs in shift-symmetric sGB theory, we numerically verified that the mode frequencies extracted from simulations is consistent with the frequencies calculated with the perturbative method~\cite{Chung:2024vaf}. We also observe the splitting of mode frequencies for the polar and axial modes in this theory. However, we find that for astrophysically realistic system, e.g. equal-mass BBH merger, and if we further limit ourselves to theory parameters where the weak coupling condition is respected and the evolution is hyperbolic, the mode splitting at the ringdown stage might be hard to extract from numerical simulations or measured from real observations due to the extremely close mode frequencies and the damping nature of QNMs. The situation will get worse for unequal mass cases as the largest coupling allowed is determined by the smaller BH.

We performed a series of nearly equal-mass, non-spinning BBH merger simulations in shift-symmetric sGB gravity with different coupling constants. We systematically extracted the QNM amplitudes and phases. Although the largest coupling used in our simulations is close to the limit given by the loss of hyperbolicity and already outside the weak coupling regime where the sGB gravity can be regarded as an EFT, the largest relative deviation for mode excitation is only around $10\%$ for the modes we extracted while for the dominant $220$ mode the change is only around $2\%$. In frequency-agnostic fitting, we find that the deviation of mode frequencies is roughly proportional to $\lambda^2$, which is consistent with the leading order theoretical predictions~\cite{Chung:2024vaf}. We note that the theoretical perturbative calculations of higher overtone frequencies are important for extracting them from the numerical simulations due to the short damping time scale of these modes. Since such overtone calculations are lacking for the moment, we could not reliably extract them from our simulations. We perform eccentricity reduction for one of our simulations to estimate the effect of the eccentricity caused by the initial growth of the scalar field. The results show that the influence of eccentricity is weaker compared to the amplitude changes caused by the presence of a scalar field. We also tried to extract the scalar-led mode in our simulations. However, only in the single BH perturbation simulations we see some evidence of the presence of such modes. For BBH merger, we are not able to reliably extract the scalar-led mode using current procedure and simulations.

Besides the shift-symmetric sGB gravity, we also consider the sGB gravity with quadratic coupling that admits the phenomena of spin-induced scalarization and dynamical scalarization. For the spin-induced scalarization case, the BBH merger only deviates from GR at the ringdown stage. By fine-tuning the parameter, we give the largest possible deviation of the ringdown waveform from GR. The relative deviation in the waveform amplitude is rather small and only at $1\%$ level, which is similar to the dominant mode in shift-symmetric case. 
We also note that for the considered models with larger coupling, which admits a faster growth of the scalar field, the maximum scalar field we could obtain, and thus the maximum deviations from GR, were relatively small. This constraints come from the loss of hyperbolicity. On the other hand, a smaller coupling, for which hyperbolic evolution is possible for relatively larger scalar fields,  leads to a much longer growth time of the scalar field. Thus, the ringdown is already decayed when the scalar field is developed.

For ringdown observations of the dynamical scalarization case, as the QNM frequencies are the same as in GR, one can only constrain the theory with ringdown mode amplitude. We perform equal-mass, non-spinning BBH simulations in the specific parameter space that allows for  dynamical scalarization. However, similar to the other cases, the change in the ringdown mode amplitude is only at a few percent level. This might cause difficulties in distinguishing this theory from GR from ringdown observations. 

Our work first numerically verified the QNM frequencies calculated with perturbative methods in shift-symmetric sGB gravity. On the other hand, the extraction of QNM amplitudes and phases can be a valuable input for building surrogate waveform models. Though our study suggests that the deviation from GR at the ringdown stage is rather small for sGB gravity, even when the coupling is close to the limit of loss of hyperbolicity and out of the range of weak coupling, future observations with full inspiral-merger-ringdown analysis can still constrain the theory.

\acknowledgments

We thank Paolo Pani for the helpful discussions.
Z.H.\ and L.S.\ are supported by the National Natural Science Foundation of China
(124B2056, 12573042), the Beijing Natural Science Foundation (1242018), the National SKA Program of China
(2020SKA0120300), and the Max Planck Partner Group Program funded by the Max Planck Society. 
Z.H.\ is supported by the China Scholarship Council (CSC). DD acknowledges financial support via an Emmy Noether Research Group funded by the German Research Foundation (DFG) under grant no. DO 1771/1-1, and the Spanish Ministry of Science and Innovation via the Ram\'on y Cajal programme (grant RYC2023-042559-I), funded by MCIN/AEI/ 10.13039/501100011033. This study is partly financed by the European Union-NextGenerationEU, through the National Recovery and Resilience Plan of the Republic of Bulgaria, project No. BG-RRP-2.004-0008-C01 and by the Spanish Agencia Estatal de Investigaci\'on (grant PID2024-159689NB-C21) funded by the Ministerio de Ciencia, Innovaci\'on y Universidades. We acknowledge Discoverer PetaSC and EuroHPC JU for awarding this project access to Discoverer supercomputer resources. We thank the entire \texttt{GRTL} Collaboration\footnote{\texttt{www.grtlcollaboration.org}} for their support and code development work. 

\appendix

\section{Eccentricity effect}\label{app:ecc}

To explore the eccentricity effects in our simulations, we perform eccentricity reduction for one of our simulations and compare the mode extraction results. We choose the simulation with $\lambda/M^2=0.05$ as it has the largest eccentricity. For performing the eccentricity extraction and reduction procedure, we set the initial distance of the two BHs to be $d=14\, M$, which is a bit further compared to the simulations analyzed in the main text. This is done in order to facilitate the eccentricity measurements, which are better performed for large distances and longer evolution. All other simulation parameters are chosen to be the same.

We estimate the eccentricity of our simulations using the method described in Refs.~\cite{Buonanno:2010yk,Knapp:2024yww}, which fits the time derivative of the orbital frequency, $\dot{\Omega}$, with a cosine function. Roughly speaking, according to Newtonian dynamics, the amplitude of the cosine function is then proportional to the orbital eccentricity. In simulations, we estimate the orbital frequency from the puncture data, which traces the motion of the two BH punctures in the coordinate. We calculate $\Omega$ using Euclidean geometry directly from the puncture coordinates, since we are only aiming to estimate the eccentricity. The distance between the two BHs at the first several orbits is relatively large, and a Newtonian approximation should be acceptable. There are methods that estimate the eccentricity from the GW waveform~\cite{Mroue:2010re}, which is a more gauge-independent way. We use the Newtonian method for the eccentricity reduction, which is discussed later.

\begin{figure}[htbp]
  \centering
  \includegraphics[width=0.45\textwidth]{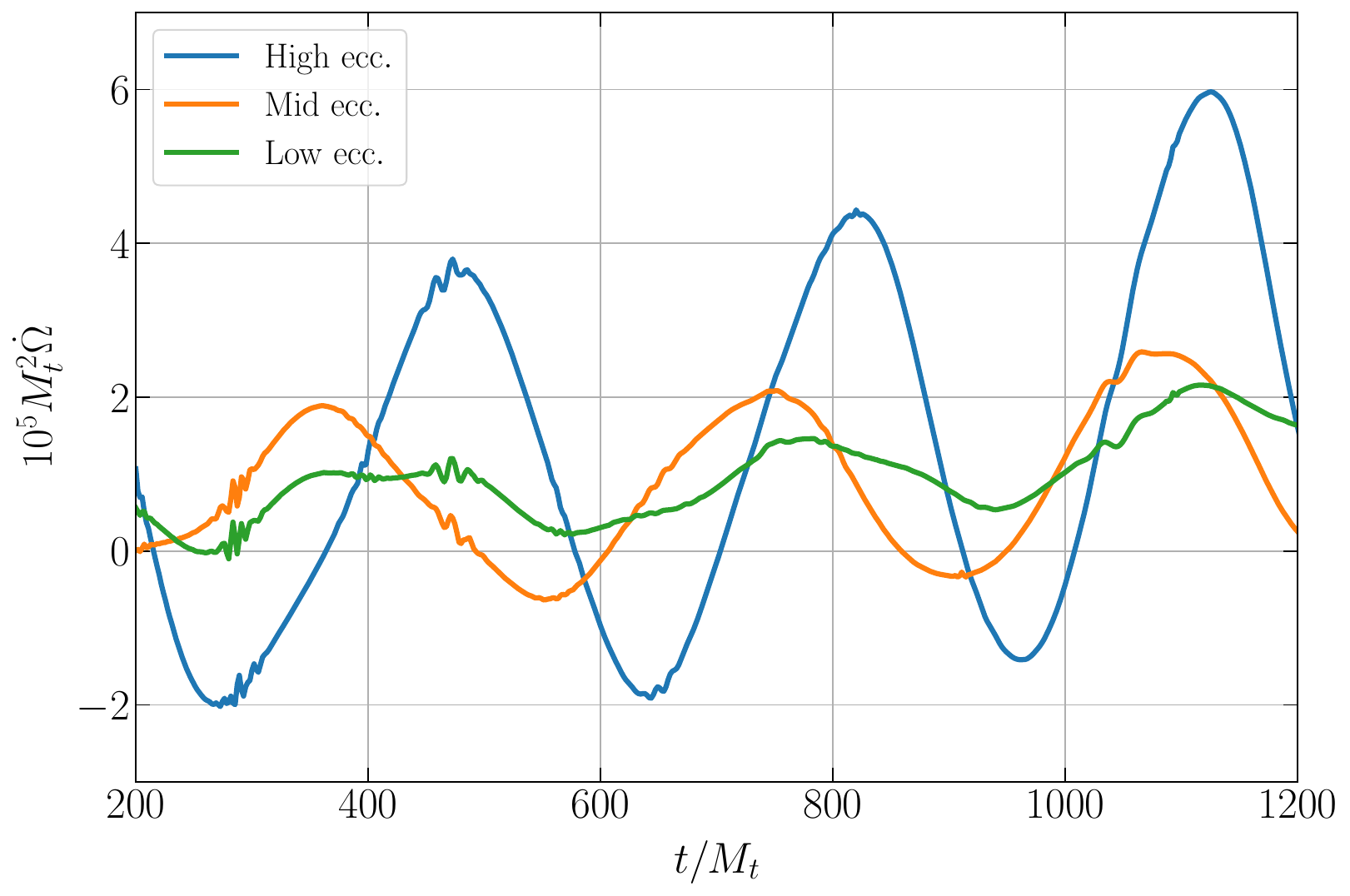}
  \caption{The time derivative of orbital frequencies, $\dot{\Omega}$, of three simulations. The simulation marked by ``High'' is the original simulation using PN initial data without eccentricity reduction, while ``Mid'' and ``Low'' mark the simulations with smaller eccentricities, obtained after performing several cycles of eccentricity reduction. \label{fig:dOmegadt}}
\end{figure}

In Fig.~\ref{fig:dOmegadt}, we show $\dot{\Omega}$ for three simulations with different eccentricities. The original simulation with the largest eccentricity is labeled ``High'' and it clearly shows the largest oscillation amplitude. The initial data of the original simulation relies on the 3PN approximation of the initial momentum of the two BHs~\cite{Healy:2017zqj,Brown:2007jx}, while the scalarization introduces additional eccentricity. The ``Mid'' and ``Low'' labels refer to two simulations with smaller eccentricities obtained by doing several cycles of eccentricity reduction.

In the following, we give a brief description of the eccentricity extraction and reduction procedure according to Ref.~\cite{Buonanno:2010yk,Knapp:2024yww}. To extract the eccentricity, we fit $\dot{\Omega}$ in a time window $t/M_t\in[500,1200]$ with 
\begin{equation}
    \dot{\Omega}(t)=S_{\Omega}(t)+B_{\Omega}\cos(\omega_{\Omega}t+\phi_{\Omega})+B_{S}\cos(2\omega_{\Omega}t+\phi_S)\,,
\end{equation}
 where $B_{\Omega}\cos(\omega_{\Omega}t+\phi_{\Omega})$ accounts for the Newtonian correction when the eccentricity is small. $B_{S}\cos(2\omega_{\Omega}t+\phi_S)$ accounts for the precession effect, which can be ignored here.
 \begin{equation}
     S_{\Omega}(t)=A_1(T_c-t)^{-11/8}+A_2(T_c-t)^{-13/8}\,,
 \end{equation}
accounts for the radiation reaction. The free parameters of the fitting are chosen to be $T_c$, $A_1$, $A_2$, $B_\Omega$, $\omega_\Omega$, $\phi_\Omega$, $B_S$ and $\phi_S$. According to Newtonian theory, when the eccentricity is small, one can write
\begin{equation}
    e=|B_\Omega|/2\omega_{\Omega}^2\,.
\end{equation}

Using the above procedure, we obtain the estimated eccentricities for the three simulations shown in Fig.~\ref{fig:dOmegadt}. From high to low, they are $e=0.037$, $0.020$, and $0.0073$ respectively. As we mentioned before, the two simulations with lower eccentricity are obtained by performing eccentricity reduction based on the original simulation. From the fitting result, one can also obtain a correction of the initial data based on Newtonian theory, which reads~\cite{Buonanno:2010yk,Knapp:2024yww}
\begin{eqnarray}\label{eq:ecc_cor}
    \Delta \dot{r}&=&\frac{r_0B_\Omega}{2\omega_{\Omega}}\cos\phi_\Omega\,,\label{eq:ecc_cor1}\\
    \Delta \Omega&=&-\frac{B_\Omega}{4\omega_{\Omega}}\sin\phi_\Omega\,,\label{eq:ecc_cor2}
\end{eqnarray}
where $\dot{r}$ and $\Omega$ are the initial relative radial velocity and orbital angular velocity of the two BHs. $\Delta\dot{r}$ and $\Delta\Omega$ are the predicted corrections. For the \texttt{TwoPunctures} solver we used, we convert these corrections to the corrections of initial puncture momentum also through Newtonian formulas. With new initial data, one can again evolve the system for several orbits and repeat the above procedure. 

\begin{table}[h]
    \centering
    \caption{The initial momentum of the puncture for the three simulations. $P_t$ is the tangential momentum while $P_r$ is the radial momentum.}\label{tab:initial data}
    \renewcommand{\arraystretch}{1.2}
    \begin{tabular}{c c c}
        \hline
         Simulation & $P_t/M$ & $P_r/M$ \\
         \hline
        High ecc. & $0.0893137945$ & $-0.0005095189$ \\
        Mid ecc. &  $0.0910142338$ & $-0.0015518286$ \\
        Low ecc. &  $0.0905$ & $-0.001$ \\
        \hline
    \end{tabular}
\end{table}

In GR simulations, the above-described eccentricity reduction procedure has been shown to be efficient~\cite{Buonanno:2010yk,Knapp:2024yww}. However, in our simulations, the eccentricity is mainly caused by the scalarization of the two BHs rather than the incorrect initial momentum. Therefore, a correction as shown in Eq.~(\ref{eq:ecc_cor1}) and Eq.~(\ref{eq:ecc_cor2}) does not give an ideal prediction of the ``correct'' initial data we need. With two iterations of the eccentricity reduction procedure, we obtain the simulation with middle eccentricity. For further iterations, although it shows a decrease in the eccentricity, we also find that the initial data converge rather slowly, and the eccentricity corrections result in an overshoot of the true low-eccentricity initial momenta (if we assume the reduction procedure will finally converge to a point that gives the lowest eccentricity). By observing the trend of the corrections, we directly guess an initial data that gives the simulation with low eccentricity. The initial momenta of the puncture used in the initial data for the three simulations are listed in Table~\ref{tab:initial data}.

\begin{figure}[htbp]
  \centering
  \includegraphics[width=0.45\textwidth]{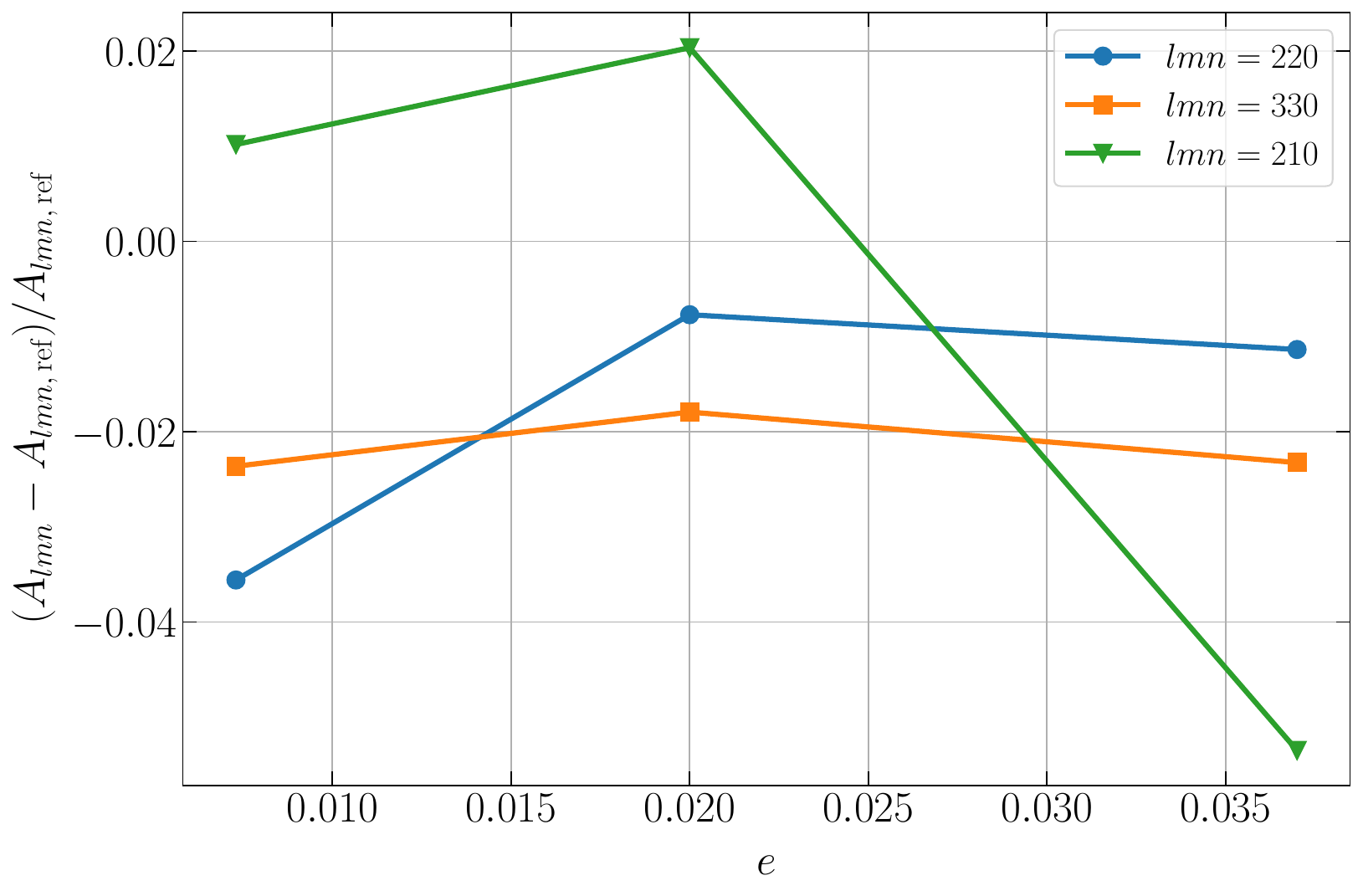}
  \caption{The relative change of the $220$, $330$, and $210$ mode amplitudes for the three simulations in Table \ref{tab:initial data}. $A_{lmn}$ is the mode amplitude and $A_{lmn,\,{\rm ref}}$ is the reference mode amplitude that refers to the amplitude that we extract in Sec.~\ref{sec:BBH} for the $\lambda/M^2=0.05$ simulation there. Note that the simulation in Sec.~\ref{sec:BBH} is different from the simulation with the largest eccentricity here, as they have different initial BH separation. \label{fig:A_e_3modes}}
\end{figure}

With these three simulations having different eccentricities, we extract the $220$, $330$, and $210$ mode frequencies based on the process discussed in Sec.~\ref{sec:extract procedure} and Sec.~\ref{sec:BBH}. In Fig.~\ref{fig:A_e_3modes} we show the relative change of the $220$, $330$, and $210$ mode amplitudes as functions of eccentricity. The reference amplitude $A_{lmn,\,{\rm ref}}$ is the amplitude extracted in Sec.~\ref{sec:BBH} for the $\lambda/M^2=0.05$ simulation there. Note that the simulations in Sec.~\ref{sec:BBH} differs from the simulations here as they have different initial BH separations. Nevertheless, Fig.~\ref{fig:A_e_3modes} can guide us on how the amplitudes calculated in Sec.~\ref{sec:BBH} change compared to more proper simulations, i.e., with larger initial separation and lower eccentricity. Despite the extraction uncertainties, we can see that there is no clear trend in mode amplitudes caused by the initial eccentricity. It seems that the simulation analyzed in Sec.~\ref{sec:BBH} gives an overall larger excitation compared to the simulations we performed here at $2\%$ level. However, this does not alter our conclusion in the main text as the change is smaller than the change caused by $\lambda$. More importantly, if we compare the three simulations performed here with the same initial separation, they only show a very small effect of eccentricity.

\bibliography{refs.bib}

\end{document}